\documentclass[aps,prd,twocolumn]{revtex4-1}
\usepackage{amsfonts}
\usepackage{amssymb}
\usepackage{amsmath}
\usepackage{color}
\usepackage[usenames,dvipsnames]{xcolor}
\usepackage{float}
\usepackage{accents}
\usepackage{graphicx}
\usepackage{epstopdf}
\usepackage{soul}
\usepackage[colorlinks=true,pdfstartview=FitV,linkcolor=blue,citecolor=blue,urlcolor=blue,breaklinks=true]{hyperref}

\begin{document}

\title{BPS Maxwell-Chern-Simons vortices with internal structures: the
Abelian Higgs and the gauged $CP(2)$ cases}
\author{J. Andrade}\email{joao.luis@discente.ufma.com.br}
\author{Rodolfo Casana} \email{rodolfo.casana@ufma.br}\email{rodolfo.casana@gmail.com}
\affiliation{Departamento de F\'{\i}sica, Universidade Federal do
Maranh\~{a}o, {65080-805}, S\~{a}o Lu\'{\i}s, Maranh\~{a}o, Brazil.} 
\author{E. da Hora}\email{carlos.hora@ufma.br}\email{edahora.ufma@gmail.com }
\affiliation{Coordenadoria Interdisciplinar de Ci\^{e}ncia e Tecnologia,
Universidade Federal do Maranh\~{a}o, {65080-805}, S\~{a}o Lu\'{\i}s, Maranh\~{a}o, Brazil.}

\begin{abstract}
We investigate the existence of first-order vortices inherent to both the
Maxwell-Chern-Simons-Higgs and the Maxwell-Chern-Simons-$CP(2)$ models
extended via the inclusion of an extra scalar sector which plays the role of
a source field. For both cases, we focus our attention on the
time-independent configurations with radial symmetry which can be obtained
through the implementation of the so-called Bogomol'nyi-Prasad-Sommerfield
(BPS) prescription. In this sense, in order to solve the corresponding
first-order differential equations, we introduce some particular scenarios
which are driven by the source field whose presence, we expect, must change
the way the resulting vortices behave. After solving the effective
first-order system through a finite-difference algorithm, we comment about
the main new effects induced by the presence of the source field in the
shape of the final configurations.
\end{abstract}

\maketitle

\section{Introduction}

\label{Intro}

In the context of classical field theories, the configurations with
nontrivial topology are solutions of highly nonlinear second-order field
equations driven by the presence of a symmetry-breaking potential \cite{n5}
which characterizes a phase transition. However, under exceptional
circumstances, they can also be studied as the solutions of a set of coupled
first-order differential equations obtained from the minimization of the
effective energy functional through the Bogomolnyi-Prasad-Sommerfield (BPS)
formalism \cite{n4}. Other methods which also lead to the first-order
equations include the study of the conservation of the energy-momentum
tensor \cite{ano} and the so-called On-Shell Method \cite{onshell}. In this
scenario, first-order vortices were obtained not only in the Maxwell-Higgs
model \cite{n1} but also in the Chern-Simons-Higgs \cite{cshv} and in the
Maxwell-Chern-Simons-Higgs \cite{mcshv} ones.

More recently, based on the exploration of the phenomenological relation
between the gauged $CP(N)$ and the 4D Yang-Mills models, Loginov \cite%
{loginov} has shown the existence of vortex solutions in a Maxwell-$CP(2)$
model. By following such an approach, some of us have obtained first-order
vortices in a Maxwell-$CP(2)$ \cite{casana}, in a Chern-Simons-$CP(2)$ \cite%
{vini} and also in a Maxwell-Chern-Simons-$CP(2)$ \cite{neyver} theories.

Another interesting issue is the existence of first-order vortices in the
context of enlarged gauged models. In this context, it is worthwhile to
highlight that the usual Maxwell-Higgs model itself was extended to
accommodate an $SO(3)$ group by the inclusion of an extra scalar sector \cite%
{witten}. In addition, a recent work \cite{n44} has verified that such
enlarged theory indeed supports first-order vortices with \textit{internal
structures}, which may find relevant applications in the context of
metamaterials \cite{n45}. Besides this point, the existence of such vortices
was also shown to occur in an extended Maxwell-$CP(2)$ model \cite{joao}.

The manuscript aims to go further by investigating whether first-order vortices with internal structures can be obtained in both Maxwell-Chern-Simons-Higgs and Maxwell-Chern-Simons-$CP(2)$ when enlarged to include an extra real scalar field.

We have organized our results as follows: In Sec. II, we define an extended Maxwell-Chern-Simons-Higgs (MCSH) model via the inclusion of a real scalar field (i.e., the source field $\chi $). Besides, the extended model also contains two arbitrary functions: the superpotential $W(\chi )$ characterizing the vacuum manifold related to the source field, and a dielectric function $G(\chi )$ multiplying the Maxwell term.  We focus the attention in those time-independent configurations possessing radial symmetry obtained via the Bogomol'nyi-Prasad-Sommerfield (BPS) algorithm. The technique provides the Bogomol'nyi bound for the total energy and the first-order differential equations themselves. The simplified structure of the BPS equation for the source field allows us to obtain the solution by fixing the superpotential $W$. In the sequence, we solve the remaining two BPS equations and the Gauss law through a finite-difference scheme for two different choices of the dielectric function $G$. Further, we depict the numerical results for the relevant fields and identify the internal structure caused by the presence of the source field.
In the Sec. III, we study an extended Maxwell-Chern-Simons-$CP(2)$ (MCS-$CP(2)$) model within the context presented in the previous section but more complicated from the technical point-of-view. In both extended models, the solutions share similar effects produced by the dielectric function. We end our manuscript in Sec. IV, in which we summarize our results and enunciate our perspectives regarding future investigations.

In what follows, we use $\eta ^{\mu \nu }=(+--)$ as the metric signature of
the (2+1)-dimensional flat spacetime, together with the natural units
system, for simplicity.

%%%%%%%%%%%%%%%%%%%%%%%%

\section{The Maxwell-Chern-Simons-Higgs case \label{2}}

In this Section, we begin the investigation about the existence of
first-order vortices with internal structures in an extended
Maxwell-Chern-Simons-Higgs scenario. Here, it is worthwhile to emphasize
some useful observations. The first one is that in typical MCSH scenario,
the correct implementation of the BPS prescription is known to be possible
only when the original model is enlarged via the inclusion of a neutral
scalar field $\Psi $ \cite{mcshv}. Second, the occurrence of the internal
structures depends on the presence of an \textit{additional} real scalar
field $\chi $ (the so-called \textit{source field}) \cite{n44}.

Therefore, based on the positive results concerning the insertion of
internal structures in the Maxwell-Higgs \cite{n44} and in the Maxwell-$%
CP(2) $ \cite{joao} models, we propose as our starting-point the following
Lagrangian density in order to study vortices with internal structure in a
Maxwell-Chern-Simons-Higgs scenario (here, $\kappa $ and $e$ are the usual
coupling constants, while $v$ stands for the vacuum expectation value for
the complex Higgs field),%
\begin{eqnarray}
\mathcal{L} &=&-\frac{G(\chi )}{4}F_{\mu \nu }F^{\mu \nu }-\frac{\kappa }{4}%
\epsilon ^{\mu \nu \kappa }A_{\mu }F_{\nu \kappa }+D^{\mu }\phi \overline{%
D_{\mu }\phi }  \notag \\[0.2cm]
&&+\frac{G(\chi )}{2}\partial _{\mu }\Psi \partial ^{\mu }\Psi +\frac{1}{2}%
\partial _{\mu }\chi \partial ^{\mu }\chi -U\left( \left\vert \phi
\right\vert ,\Psi ,\chi \right) \text{,}  \label{1}
\end{eqnarray}%
\textbf{\ }where $A_{\mu }$ is the electromagnetic field, $F_{\mu \nu
}=\partial _{\mu }A_{\nu }-\partial _{\nu }A_{\mu }$ represents its strength
tensor and $\epsilon ^{\mu \nu \kappa }$ is the Levi-Civita tensor (with $%
\epsilon ^{012}=+1$). The coupling between the electromagnetic and Higgs
fields is accounted by the standard covariant derivative,%
\begin{equation}
D_{\mu }\phi =\partial _{\mu }\phi -ieA_{\mu }\phi \text{.}
\end{equation}%
We also point out the presence of a dielectric function $G\equiv G(\chi )$
multiplying both the Maxwell and the kinetic term for the field $\Psi $. As
we demonstrate below, such a function is the responsible for the formation
of the internal structure inherent to the vortex configurations.

In the present case, we choose the potential as%
\begin{eqnarray}
U(|\phi |,\Psi ,\chi ) &=&\frac{1}{2G}\left[ e\left( v^{2}-\left\vert \phi
\right\vert ^{2}\right) -\kappa \Psi \right] ^{2}  \notag \\[0.2cm]
&&+e^{2}\left\vert \phi \right\vert ^{2}\Psi ^{2}+\frac{1}{2r^{2}}\left(
\frac{dW}{d\chi }\right) ^{2}\text{,}  \label{p}
\end{eqnarray}%
where the first two terms in the right-hand side (with $G=1$) correspond to
the potential which allows the standard Maxwell-Chern-Simons-Higgs model to
be self-dual \cite{mcshv}. In our enlarged case, the third term stands for
the potential related to the source field, with $W=W(\chi )$ being called
\textit{superpotential}. Here, even though the breaking of the translational
invariance (from which we conclude that the resulting model must be seen
therefore as an effective one), the factor $1/r^{2}$ is included to support
the obtainment of the first-order equation for the source field itself \cite%
{n44}.

The presence of the Chern-Simons term does not allow the implementation of
the gauge choice $A_{0}=0$ given that this choice does not solve the
corresponding Gauss law, from which we conclude that the resulting
structures possess nontrivial profiles for both the electric and the
magnetic fields. Consequently, it is reasonable to expect that the internal
structure will be present in both the electric and the magnetic sectors.

We look for time-independent configurations via the standard radially
symmetric map which is given by (the Latin indexes run over the spatial
coordinates only)%
\begin{equation}
\phi =vg(r)e^{in\theta }\text{ \ and \ }A_{i}=\varepsilon _{ij}\frac{x_{j}}{%
er^{2}}\left[ a(r)-n\right] \text{,}
\end{equation}%
where $r$ and $\theta $ are the polar coordinates, $\varepsilon _{ij}$
stands for the antisymmetric symbol, $x_{j}$ is the unit vector and the
integer parameter $n=\pm 1,$ $\pm 2,$ $\pm 3...$ represents the winding
number (vorticity) of the final solutions. In view of this map, we also get
that $A_{0}=A_{0}(r)$, $\Psi =\Psi (r)$ and $\chi =\chi (r)$. Moreover, the
dimensionless profile functions $g(r)$ and $a(r)$ are supposed to behave as%
\begin{equation}
g(r=0)=0\text{ \ and \ }a(r=0)=n\text{,}  \label{b1}
\end{equation}%
\begin{equation}
g(r\rightarrow \infty )\rightarrow 1\text{ \ and \ }a(r\rightarrow \infty
)\rightarrow 0\text{,}  \label{b2}
\end{equation}%
from which we expect to obtain regular solutions with finite energy.

We focus our attention on those vortex solutions satisfying a particular set
of coupled first-order differential equations. These equations arise through
the implementation of the BPS prescription (i.e. the minimization of the
energy of the model (\ref{1})). The starting-point is the radially symmetric
expression for the stationary energy distribution given by%
\begin{eqnarray}
\varepsilon &=&\frac{G}{2}B^{2}+\frac{G}{2}\left( \frac{dA_{0}}{dr}\right)
^{2}+v^{2}\left( \left( \frac{dg}{dr}\right) ^{2}+\frac{a^{2}g^{2}}{r^{2}}%
\right)  \notag \\
&&+e^{2}v^{2}g^{2}\left( A_{0}\right) ^{2}+\frac{G}{2}\left( \frac{d\Psi }{dr%
}\right) ^{2}+\frac{1}{2}\left( \frac{d\chi }{dr}\right) ^{2}  \notag \\
&&+\frac{1}{2G}\left[ ev^{2}\left( 1-g^{2}\right) -\kappa \Psi \right] ^{2}
\notag \\
&&+e^{2}v^{2}g^{2}\Psi ^{2}+\frac{1}{2r^{2}}\left( \frac{dW}{d\chi }\right)
^{2}\text{,}  \label{ed}
\end{eqnarray}%
which, after some algebraic manipulation, can be promptly rearranged as%
\begin{eqnarray}
\varepsilon &=&\varepsilon _{{BPS}}+\frac{1}{2G}\left( \frac{{}}{{}}GB\mp %
\left[ ev^{2}\left( 1-g^{2}\right) -\kappa \Psi \right] \right) ^{2}  \notag
\\
&&+v^{2}\left( \frac{dg}{dr}\mp \frac{ag}{r}\right) ^{2}+\frac{1}{2}\left(
\frac{d\chi }{dr}\mp \frac{1}{r}\frac{dW}{d\chi }\right) ^{2}  \notag \\
&&+\frac{1}{2}G\left( \frac{dA_{0}}{dr}\mp \frac{d\Psi }{dr}\right)
^{2}+e^{2}v^{2}g^{2}\left( A_{0}\mp \Psi \right) ^{2}\text{,}
\end{eqnarray}%
where we have defined the term $\varepsilon _{{BPS}}$ as%
\begin{equation}
\varepsilon _{{BPS}}=\mp \frac{v^{2}}{r}\frac{d}{dr}\left[ a\left(
1-g^{2}\right) \right] \pm \frac{1}{r}\frac{dW}{dr}\text{,}  \label{ed0}
\end{equation}%
whose integration over the plane, using the boundary conditions (\ref{b1})
and (\ref{b2}), provides the expression which becomes the Bogomol'nyi bound
for the extended MCSH model, i.e.%
\begin{equation}
E_{BPS}=2\pi \int_{0}^{\infty }r\varepsilon _{{BPS}}dr=\pm 2\pi \left(
v^{2}n+\Delta W\right) \text{,}  \label{te1}
\end{equation}%
with $\Delta W=W\left( r\rightarrow \infty \right) -W(r=0)$. In the
expression above, the upper (lower) sign holds for positive (negative)
values of both $n$ and $\Delta W$.

Considering these results, one gets that the total energy of the model
satisfies%
\begin{eqnarray}
&&\left. \frac{E}{2\pi }=\int_{0}^{\infty }\frac{1}{2G}\left( \frac{{}}{{}}%
GB\mp \left[ ev^{2}\left( 1-g^{2}\right) -\kappa \Psi \right] \right)
^{2}rdr\right.  \notag \\
&&\left. +\int_{0}^{\infty }\left[ v^{2}\left( \frac{dg}{dr}\mp \frac{ag}{r}%
\right) ^{2}+\frac{1}{2}\left( \frac{d\chi }{dr}\mp \frac{1}{r}\frac{dW}{%
d\chi }\right) ^{2}\right] rdr\right.  \notag \\
&&\left. +\int_{0}^{\infty }\left[ \frac{G}{2}\left( \frac{dA_{0}}{dr}\mp
\frac{d\Psi }{dr}\right) ^{2}+e^{2}v^{2}g^{2}\left( A_{0}\mp \Psi \right)
^{2}\right] rdr\right.  \notag \\
&&\left. +\frac{E_{BPS}}{2\pi }\geq v^{2}\left\vert n\right\vert +\left\vert
\Delta W\right\vert \text{,}\right.
\end{eqnarray}%
which reveals that the energy is indeed bounded from below and that the
bound is saturated when the fields satisfy%
\begin{equation}
GB=\pm ev^{2}\left( 1-g^{2}\right) \mp \kappa \Psi \text{,}  \label{bps1}
\end{equation}%
\begin{equation}
\frac{dg}{dr}=\pm \frac{ag}{r}\text{,}  \label{bps2}
\end{equation}%
\begin{equation}
\frac{d\chi }{dr}=\pm \frac{1}{r}\frac{dW}{d\chi }\text{,}  \label{sf}
\end{equation}%
\begin{equation}
\frac{d\Psi }{dr}=\pm \frac{dA_{0}}{dr}\text{ \ and \ }\Psi =\pm A_{0}\text{.%
}  \label{BPS44}
\end{equation}%
Therefore, the system above stands for the first-order BPS equations of the
model (also here, the upper (lower) sign holds for $n>0$ ($n<0$)). In
particular, when these first-order equations are satisfied, the energy of
the resulting BPS configurations is equal to $E_{BPS}$, the corresponding
Bogomol'nyi bound, calculated in (\ref{te1}).

In view of the Eq. (\ref{BPS44}), we point out that $\Psi =\pm A_{0}$ also
solves the $\Psi ^{\prime }=\pm A_{0}^{\prime }$ (here, prime denotes the
derivative with respect to the radial coordinate). As usual, the solution
for the scalar potential $A_{0}(r)$ is obtained by solving the Gauss law
which, given the radially symmetric scenario, takes the form%
\begin{equation}
\frac{1}{r}\frac{d}{dr}\left( rG\frac{dA_{0}}{dr}\right) +\kappa
B=2e^{2}v^{2}g^{2}A_{0}\text{.}  \label{gl1}
\end{equation}%
It must be solved (together with the remaining three first-order equations
above) according to the boundary conditions%
\begin{equation}
A_{0}(r=0)=w_{0}\text{ \ and \ }A_{0}^{\prime }(r\rightarrow \infty
)\rightarrow 0\text{,}  \label{b3}
\end{equation}%
where $w_{0}=A_{0}(0)$ stands for a constant.

\subsection{Some MCSH scenarios with internal structure\label{MCSHsc}}

We now proceed with the construction of some particular scenarios and their
respective numerical solutions. We begin this task by pointing out that the
first-order equation for the source field $\chi $ does not contain the other
fields of the model and depends only on the form of the superpotential $%
W(\chi )$. Here, we choose the superpotential as%
\begin{equation}
W(\chi )=\chi -\frac{1}{3}\chi ^{3}\text{,}  \label{w}
\end{equation}%
which was used recently as an attempt to understand planar skyrmion-like
solitons \cite{2324} and the behavior of massless Dirac fermions in a
skyrmion-like background \cite{25}.

In view of this choice, the first-order equation (\ref{sf}) for the source
field can be written in the form%
\begin{equation}
\frac{d\chi }{dr}=\pm \frac{1}{r}\left( 1-\chi ^{2}\right) \text{,}
\label{w1}
\end{equation}%
whose exact solution reads%
\begin{equation}
\chi (r) =\pm \frac{r^{2}-r_{0}^{2}}{r^{2}+r_{0}^{2}}\text{,}  \label{ssf}
\end{equation}%
where $r_{0}$ represents an arbitrary positive constant such that $\chi
\left( r=r_{0}\right) =0$. We note that this solution attains the boundary
values $\chi \left( r=0\right) =\mp 1$ and $\chi \left( r\rightarrow \infty
\right) \rightarrow \pm 1$.

We also rewrite the BPS energy density (\ref{ed0}) as%
\begin{equation}
\varepsilon _{BPS}=\varepsilon _{G}+\varepsilon _{\chi }\text{,}
\label{ed1xcp}
\end{equation}%
where $\varepsilon _{G}$\ represents the contribution which refers to the
vortex with internal structure, i.e.%
\begin{equation}
\varepsilon _{G}=G\left[ B^{2}+\left( \frac{dA_{0}}{dr}\right) ^{2}\right]
+2v^{2}\left[ e^{2}g^{2}\left( A_{0}\right) ^{2}+\frac{a^{2}g^{2}}{r^{2}}%
\right] \text{,}  \label{edG}
\end{equation}%
while $\varepsilon _{\chi }$ corresponds to the contribution related
exclusively to the source field $\chi $, i.e.%
\begin{equation}
\varepsilon _{\chi }=\left( \frac{d\chi }{dr}\right) ^{2}\text{.}
\label{edX}
\end{equation}

In order to proceed with our investigation, we now need to choose a
particular expression for the dielectric function $G(\chi )$. In this
regard, for pedagogical reasons, we separate our study in two different
scenarios.
\begin{figure}[t]
\includegraphics[width=8.4cm]{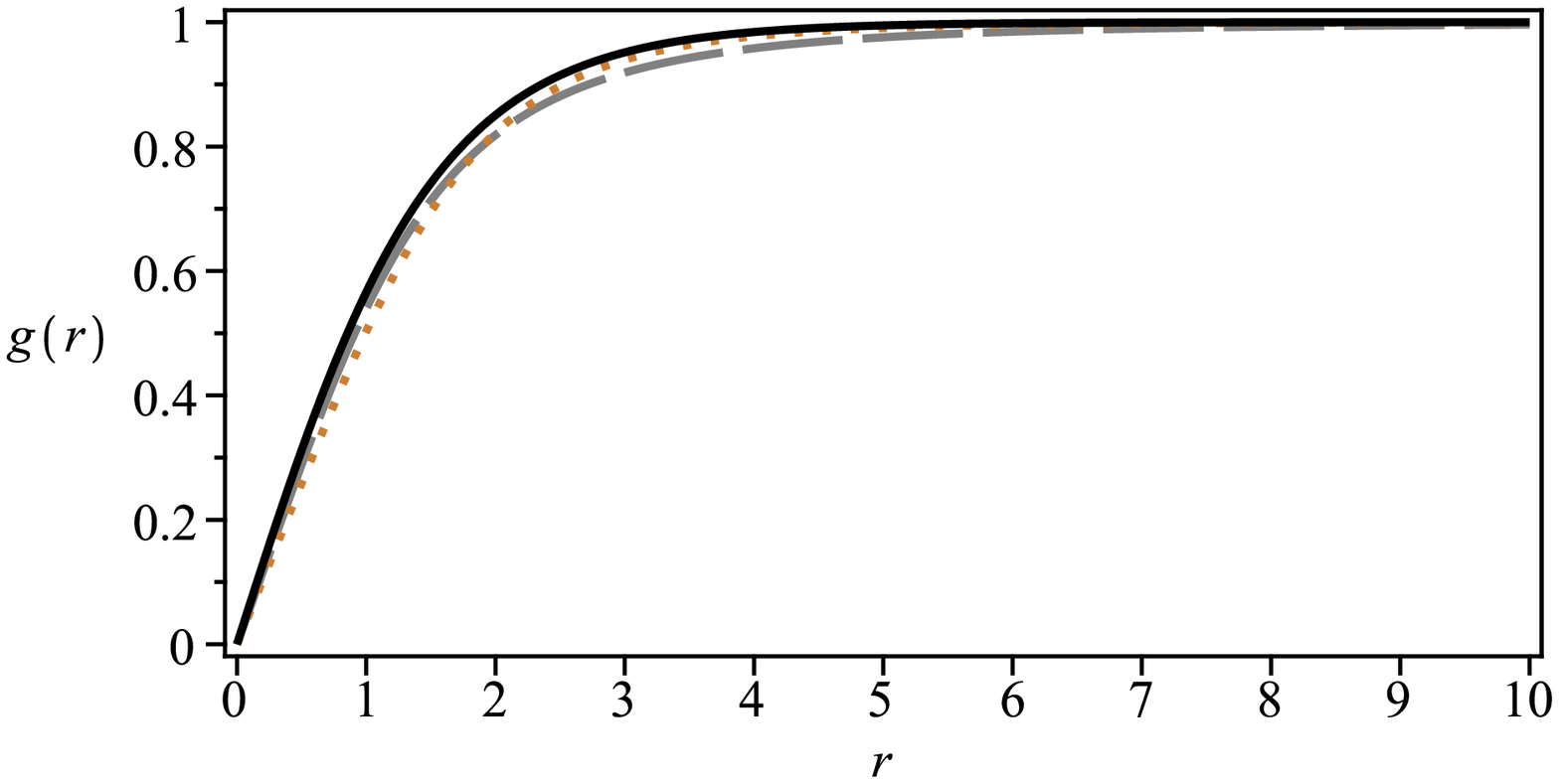} %
\includegraphics[width=8.4cm]{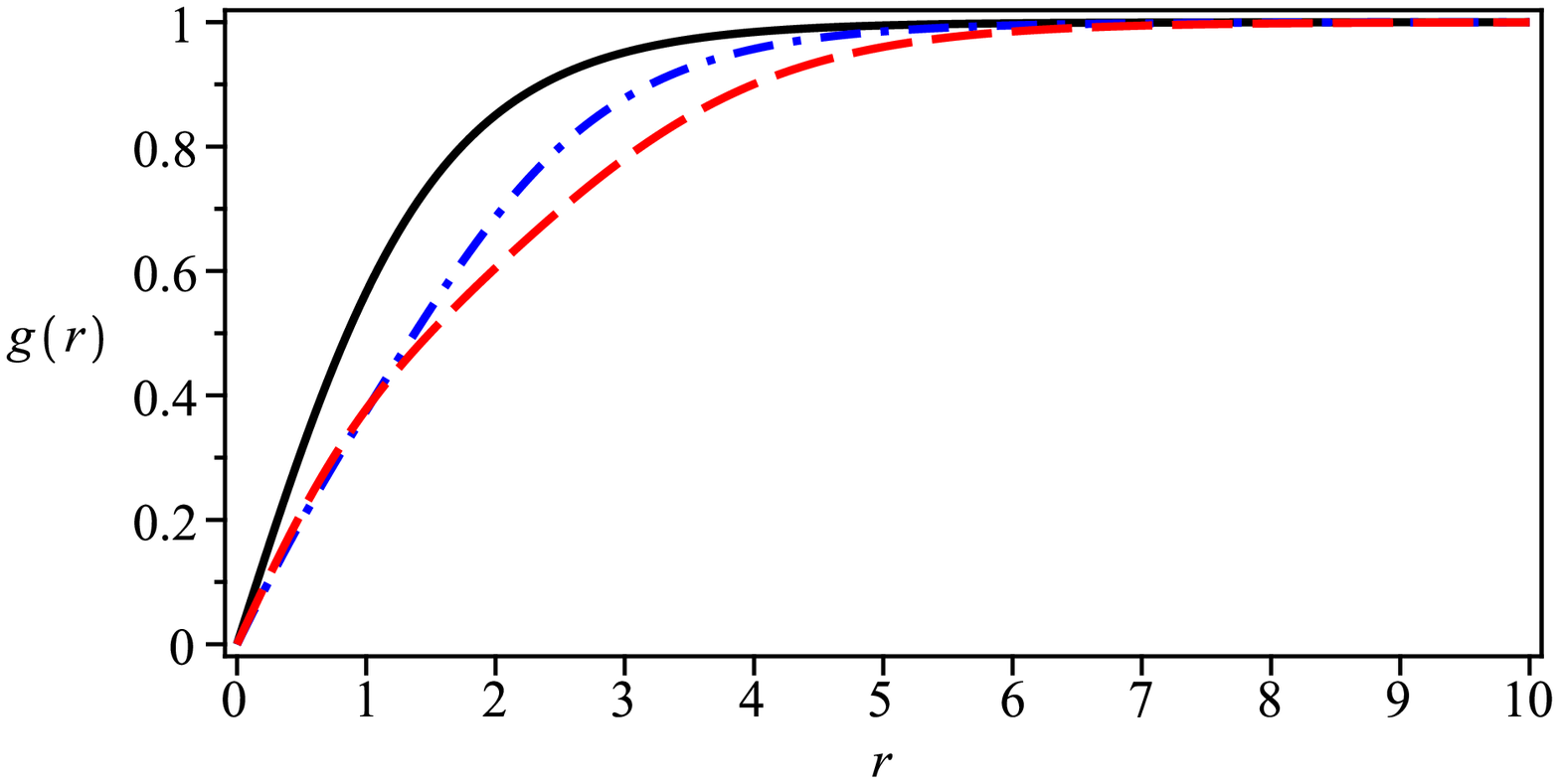}
\caption{Numerical solutions to the scalar profile function $g\left(
r\right) $ coming from (\protect\ref{b4}), (\protect\ref{b5}) and (\protect
\ref{b6}) (top, long-dashed gray line for $r_{0}=1$ and dotted orange line
for $r_{0}=2$)\ and (\protect\ref{b7}), (\protect\ref{b8}) and (\protect\ref%
{b9}) (bottom, dash-dotted blue line for $r_{0}=1$ and dashed red line for $%
r_{0}=2$). In both cases, the equations were solved according the boundary
conditions (\protect\ref{b1}), (\protect\ref{b2}) and (\protect\ref{b3}). We
have used $e=v=\protect\kappa =1$ and $n=1$\ (i.e. upper signs in the
first-order expressions). Here, the solid black line is the usual solution
(absence of the source field, with $G=1$).} \label{figg1}
\end{figure}

\subsubsection{The first case\label{fcmcsh1}}

In what follows, we choose the dielectric function as%
\begin{equation}
G(\chi )=\frac{1}{1-\chi ^{2}}\text{,}  \label{tg1}
\end{equation}%
which, taking into account the solution (\ref{ssf}), can be expressed as%
\begin{equation}
G(r)=\frac{\left( r^{2}+r_{0}^{2}\right) ^{2}}{4r^{2}r_{0}^{2}}\text{,}
\label{tg11}
\end{equation}%
which is finite along the radial axis but diverges at the boundaries (i.e.
for $r=0$ and $r\rightarrow \infty $), see the Eq. (\ref{ssf}). Therefore,
in order to avoid that the first two terms in the right-hand side of the Eq.
(\ref{ed}) to be singular, the magnetic and electric fields must vanish at
the boundaries, from which the total energy is expected to converge to the
finite value given by (\ref{te1}).

By using (\ref{tg11}), the first-order equations (\ref{bps1}) and (\ref{bps2}%
) assume the form%
\begin{equation}
\frac{1}{r}\frac{da}{dr}=\frac{4r^{2}r_{0}^{2}}{\left(
r^{2}+r_{0}^{2}\right) ^{2}}\left( \mp e^{2}v^{2}\left( 1-g^{2}\right)
+e\kappa A_{0}\right) \text{,}  \label{b4}
\end{equation}%
\begin{equation}
\frac{dg}{dr}=\pm \frac{ag}{r}\text{,}  \label{b5}
\end{equation}%
while the Gauss law can be written as%
\begin{equation}
\frac{1}{4r_{0}^{2}r}\frac{d}{dr}\left[ \frac{\left( r^{2}+r_{0}^{2}\right)
^{2}}{r}\frac{dA_{0}}{dr}\right] +\kappa B=2e^{2}v^{2}g^{2}A_{0}\text{,}
\label{b6}
\end{equation}%
where we have used%
\begin{equation}
B(r)=-\frac{1}{er}\frac{da}{dr}\text{,}  \label{mf}
\end{equation}%
which stands for the radially symmetric expression for the magnetic field.

Hereinafter, we fix the constants of the model as $e=v=\kappa =1$ and $n=1$\
(i.e. upper signs in the first-order expressions), from which we solve the
equations (\ref{b4}), (\ref{b5}) and (\ref{b6}) above for $r_{0}=1$
(long-dashed gray line) and $r_{0}=2$ (dotted orange line) via the
implementation of a finite-difference algorithm according the boundary
conditions (\ref{b1}), (\ref{b2}) and (\ref{b3}). We then depict the
numerical solutions for the profile functions $g(r)$ and $a(r)$, the
electric potential $A_{0}(r)$, the magnetic $B(r)$ and electric $E(r)$
fields, and the energy distribution $\varepsilon _{BPS}$, see the figures
1-6 below (in which the solid black line represents the usual solution).

Figures 1 and 2 show the Higgs profile $g(r)$ and the gauge one $a(r)$,
respectively. We see that, despite the presence of the source field, these
functions converge naturally to the boundary values given by (\ref{b1}) and (%
\ref{b2}). In the present case, the effects caused by the source field are
slight variations on the core-size of the corresponding solutions.

The solutions for the electric potential $A_{0}(r)$, depicted in the Fig. 3,
show that the variation of $r_{0}$ changes both the core-size and the
maximum value reached on the origin.

Figures 4 and 5 present, respectively, the profiles for the magnetic and
electric fields, which stand for ring structures. In particular, we note
that the magnetic field behavior is dramatically different from the usual
one. As we have explained previously, in order to avoid the first two terms
on the right-hand side of (\ref{ed}) to be singular, these fields vanish in
the boundaries $r=0$ and $r\rightarrow \infty $, and this fact allows the
the total energy to converge to the Bogomol'nyi bound (\ref{te1}).
Furthermore, the variations on $r_{0}$ also affect both the core-size and
the maximum value these solutions reach.
\begin{figure}[t]
\includegraphics[width=8.4cm]{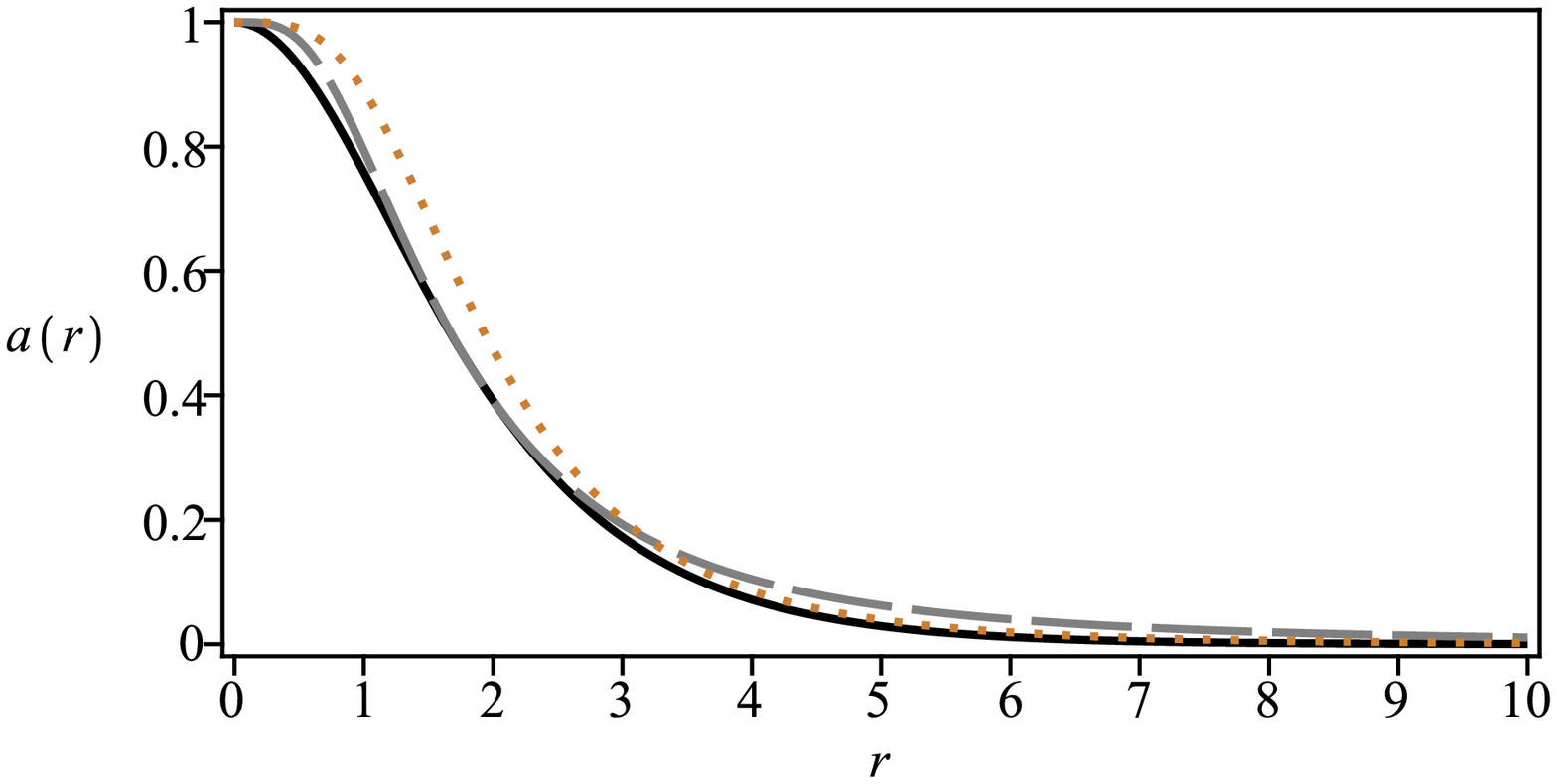} %
\includegraphics[width=8.4cm]{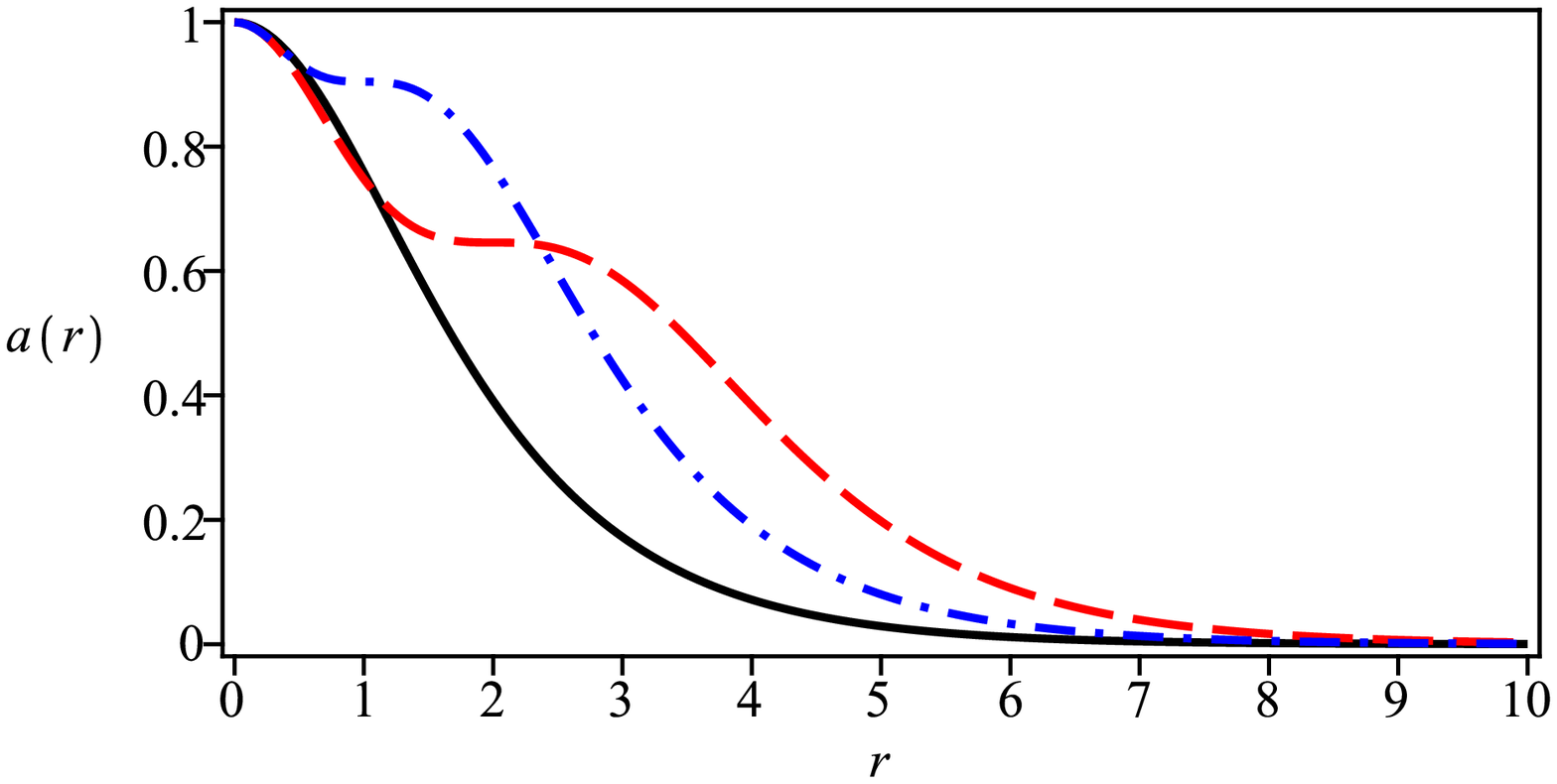}
\caption{Numerical solutions to the gauge profile function $a(r)$.
Conventions as in the Fig. \ref{figg1}. The solutions for the 1st. case mimic the
usual ones around the point $r=r_{0}$, while the profiles for the 2nd. case
have a plateau at $r=r_{0}$.}
\end{figure}

Finally, the profiles for the energy distribution $\varepsilon _{G}$ related
to the BPS configurations are depicted in the Fig. 6, from which we see how
the source field changes the shape of these solutions from a lump (standard
profile) to a ring (nonusual ones), with $r_{0}$ also controlling the
"radius" (i.e. the distance from the origin for which the solution reaches
its maximum value) of these rings.

Now, before studying a new case, it is interesting to investigate how the
dielectric medium affects the way the fundamental fields behave close to the
boundary values.\ With this purpose in mind, we solve the equations (\ref{b4}%
), (\ref{b5}) and (\ref{b6}) around the boundaries. In this case, for $n=N>0$%
, we get that, near the origin, the fields behave approximately as%
\begin{equation}
g(r)\approx g_{N}r^{N}-\frac{eB_{0}}{4r_{0}^{2}}g_{N}r^{N+4}\text{,}
\end{equation}%
\begin{equation}
a(r)\approx N-\frac{eB_{0}}{r_{0}^{2}}r^{4}\text{,}
\end{equation}%
\begin{equation}
A_{0}(r)\approx w_{0}+w_{2}r^{2}-\frac{w_{2}}{r_{0}^{2}}r^{4}\text{,}
\end{equation}%
while the solutions in the limit $r\rightarrow \infty $ read%
\begin{equation}
g(r)\approx 1-C_{\infty }r^{-\Lambda }\text{,}
\end{equation}%
\begin{equation}
a(r)\approx \Lambda C_{\infty }r^{-\Lambda }\text{,}
\end{equation}%
\begin{equation}
A_{0}(r)\approx -\frac{\Lambda C_{\infty }}{2e\kappa \left( r_{0}\right) ^{2}%
}r^{-\Lambda }\text{,}
\end{equation}%
where $g_{N}>0$, $C_{\infty }>0$ and $w_{2}$ are positive integration
constants. We have also defined%
\begin{equation}
B_{0}=B\left( r=0\right) =v^{2}e-w_{0}\kappa \text{,}  \label{B00}
\end{equation}%
i.e. the value of the magnetic field at the origin without the influence of
the dielectric medium. Moreover, we have also introduced the parameter $%
\Lambda $ as%
\begin{equation}
\Lambda =1+\sqrt{1+8\left( evr_{0}\right) ^{2}}\text{,}  \label{ww1}
\end{equation}%
which shows how the dielectric medium (via $r_{0}$) controls the way the
fields decay when $r\rightarrow \infty $.
\begin{figure}[t]
\includegraphics[width=8.4cm]{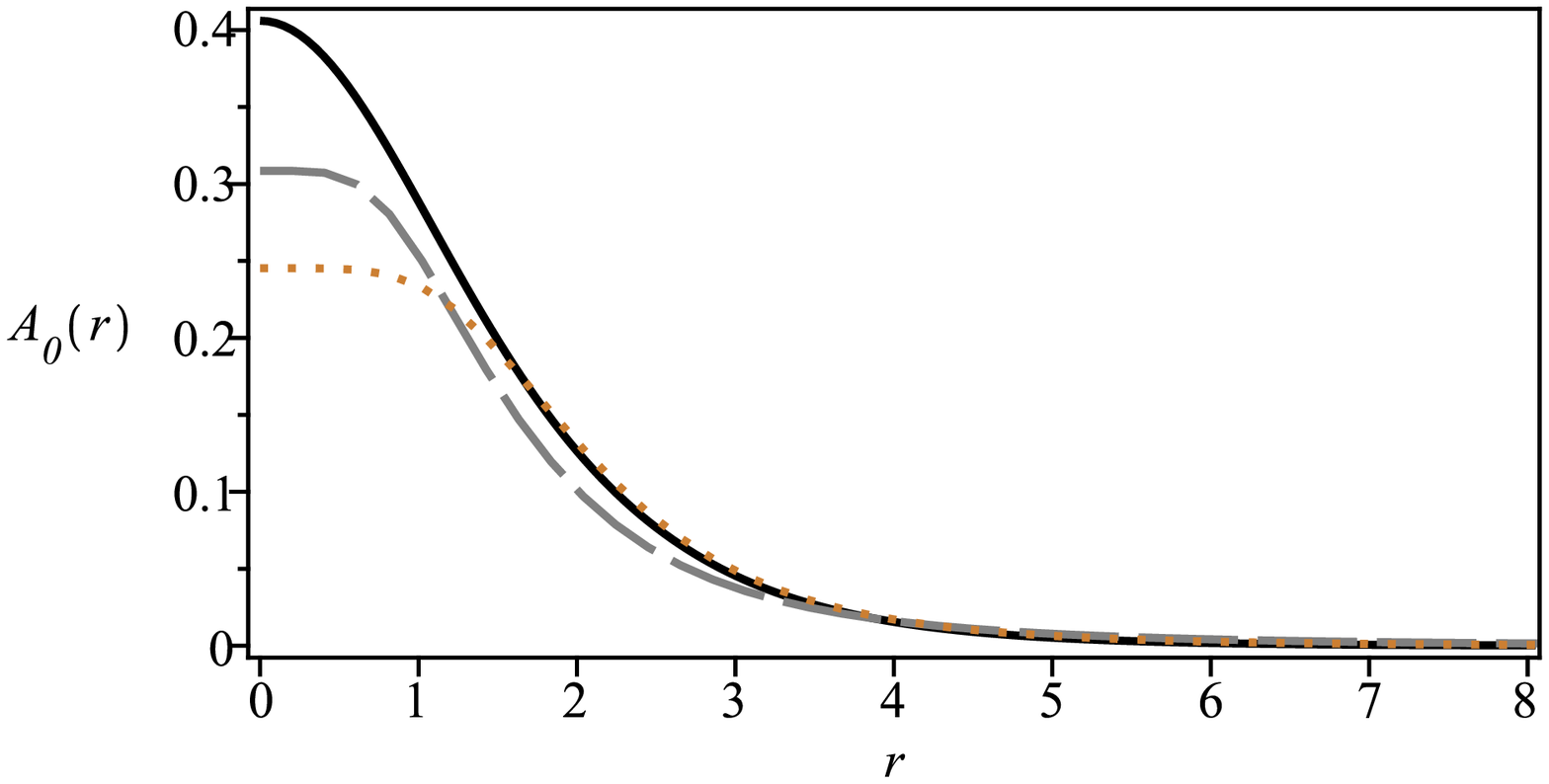} %
\includegraphics[width=8.4cm]{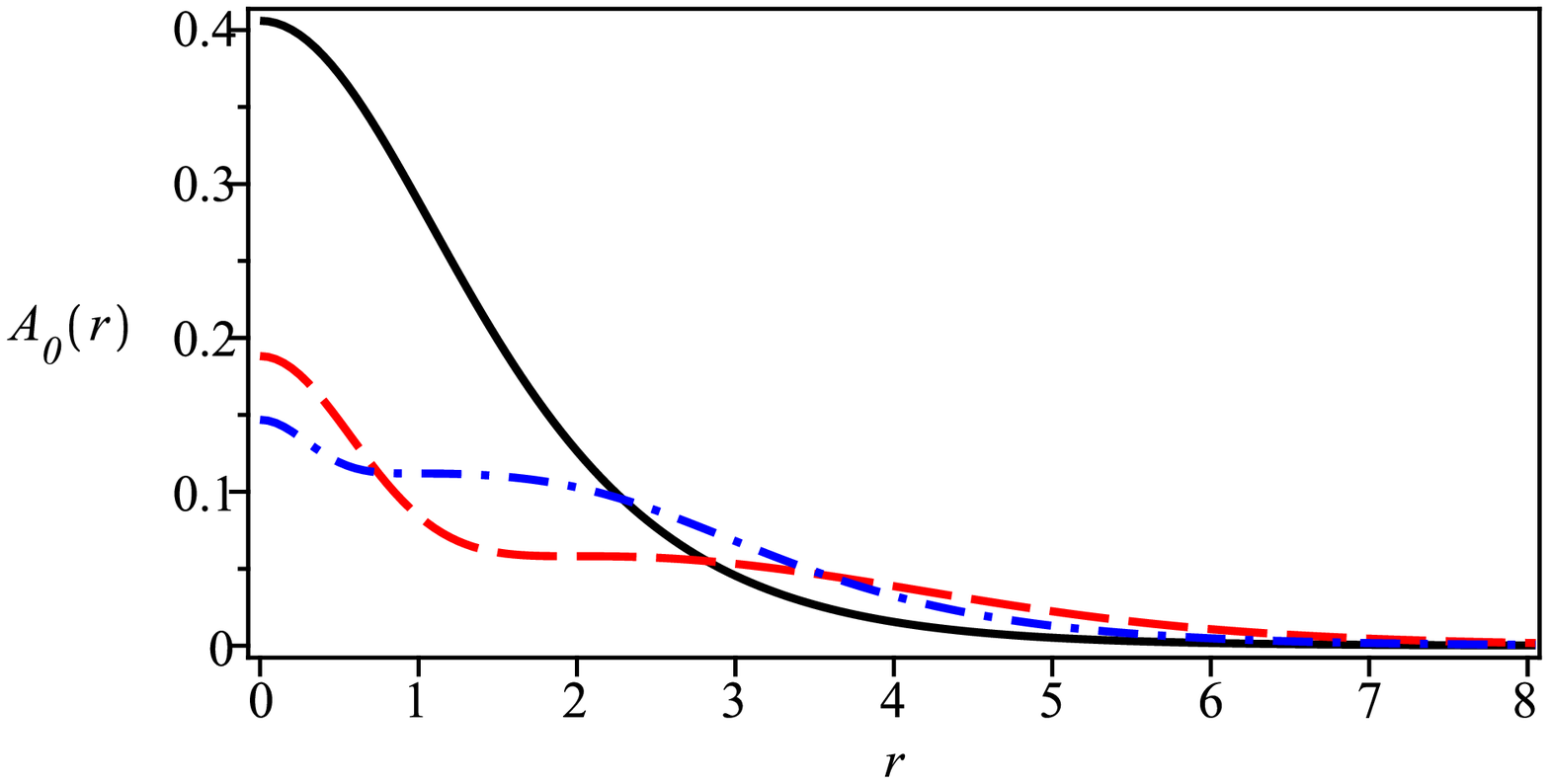}
\caption{Numerical solutions to the electric potential $A_{0}(r)$.
Conventions as in the Fig. \ref{figg1}. The solutions for the 1st. case mimic the
standard behavior around $r=r_{0}$. The results for the 2nd. case also have
a plateau at $r=r_{0}$.}
\end{figure}

\subsubsection{The second case\label{mcshcc2}}

We now consider another interesting choice for the dielectric function $%
G(\chi )$, whose expression reads%
\begin{equation}
G(\chi )=\frac{1}{\chi ^{2}}\text{,}  \label{tg2}
\end{equation}%
which behaves as $G(r=0)=1$ and $G(r\rightarrow \infty )\rightarrow 1$ when
the field $\chi $ given by the Eq.(\ref{ssf}). Consequently, at the
boundaries, the new vortex configurations\ mimic the behavior of those
obtained in the absence of the source field.
\begin{figure}[t]
\includegraphics[width=8.4cm]{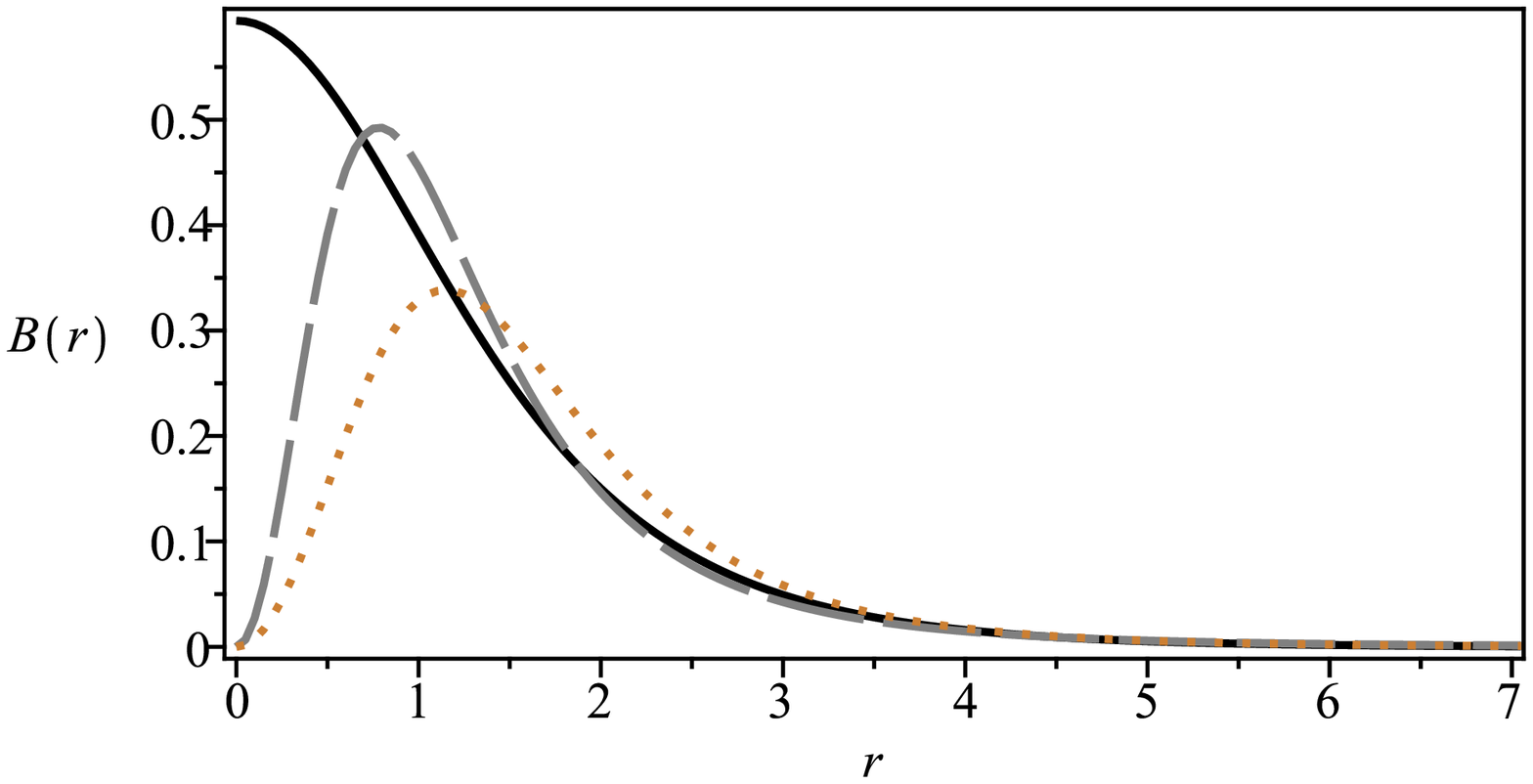} %
\includegraphics[width=8.4cm]{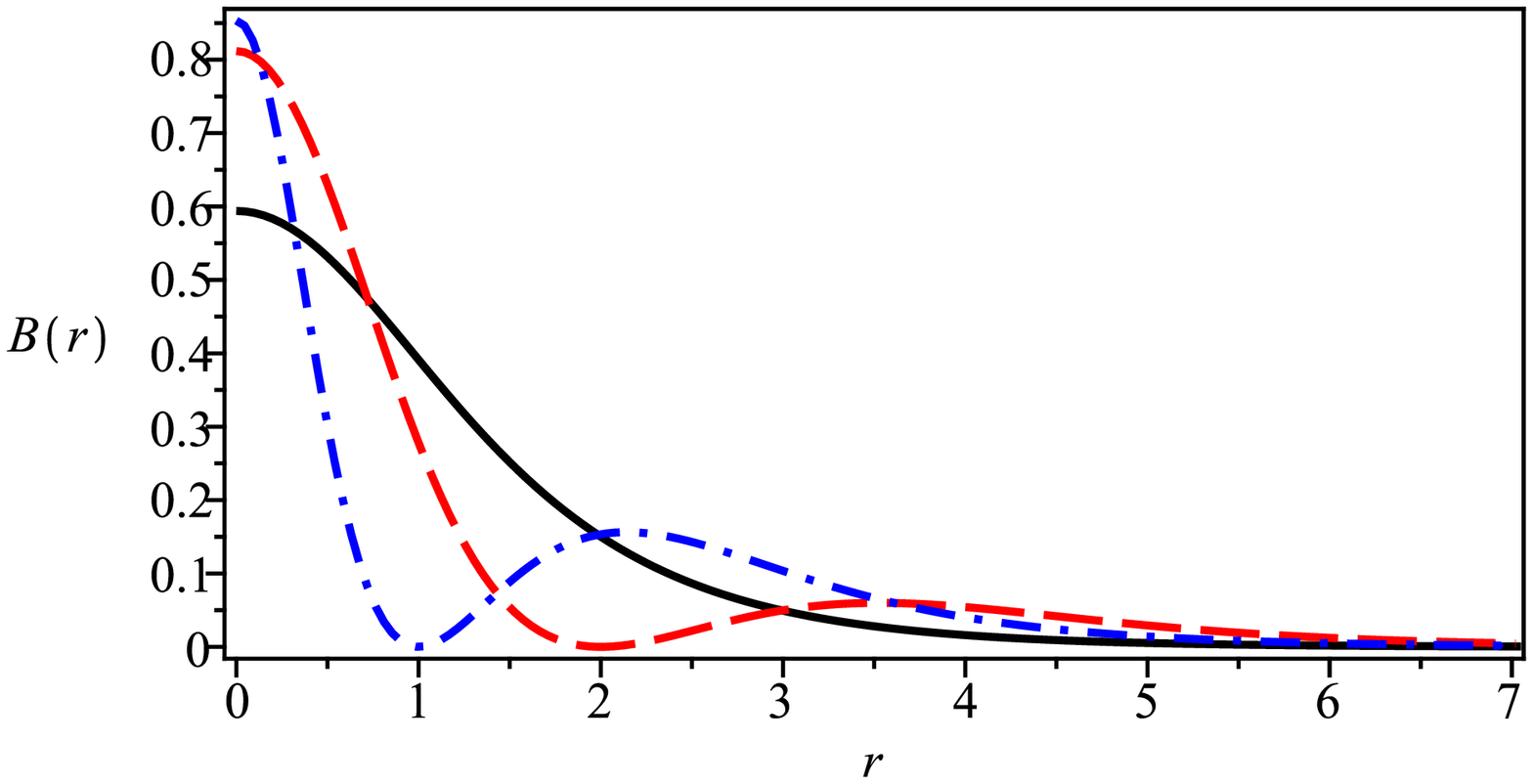}
\caption{Numerical solutions to the magnetic field $B(r)$. Conventions as in
the Fig. \ref{figg1}. The solutions for the 2nd. case vanish at $r=r_{0}$ (what can be
understood as an internal structure) in order to avoid the first term in the
right-hand side of the Eq. (\protect\ref{ed}) to be singular.}
\end{figure}

Moreover, it is interesting to highlight that, when $r=r_{0}$, the source
field $\chi $ given by (\ref{ssf}) vanishes and that therefore the function $%
G$ in (\ref{tg2}) diverges. This singularity is counterbalanced by the
electric and magnetic fields, which must vanish $r=r_{0}$. This fact leads
to a finite energy density (\ref{ed}) whose correspondent BPS total energy
converges to the value (\ref{te1}). The vanishing of the electric and
magnetic fields in $r=r_{0}$ is the manifestation of an internal structure
in the vortex configurations.

We now use the Eq. (\ref{ssf}) in order to rewrite the dielectric function
given by (\ref{tg2}) in the form%
\begin{equation}
G(r)=\frac{\left( r^{2}+r_{0}^{2}\right) ^{2}}{\left( r^{2}-r_{0}^{2}\right)
^{2}}\text{,}  \label{tg22}
\end{equation}%
which leads to the first-order equations%
\begin{equation}
\frac{1}{r}\frac{da}{dr}=\frac{\left( r^{2}-r_{0}^{2}\right) ^{2}}{\left(
r^{2}+r_{0}^{2}\right) ^{2}}\left( \mp e^{2}v^{2}\left( 1-g^{2}\right)
+e\kappa A_{0}\right) \text{,}  \label{b7}
\end{equation}%
\begin{equation}
\frac{dg}{dr}=\pm \frac{ag}{r}\text{,}  \label{b8}
\end{equation}%
and to the Gauss law%
\begin{equation}
\frac{1}{r}\frac{d}{dr}\left[ \frac{\left( r^{2}+r_{0}^{2}\right) ^{2}r}{%
\left( r^{2}-r_{0}^{2}\right) ^{2}}\frac{dA_{0}}{dr}\right] +\kappa
B=2e^{2}v^{2}g^{2}A_{0}\text{,}  \label{b9}
\end{equation}%
where we have again used the Eq. (\ref{mf}) for the magnetic field.
\begin{figure}[t]
\includegraphics[width=8.4cm]{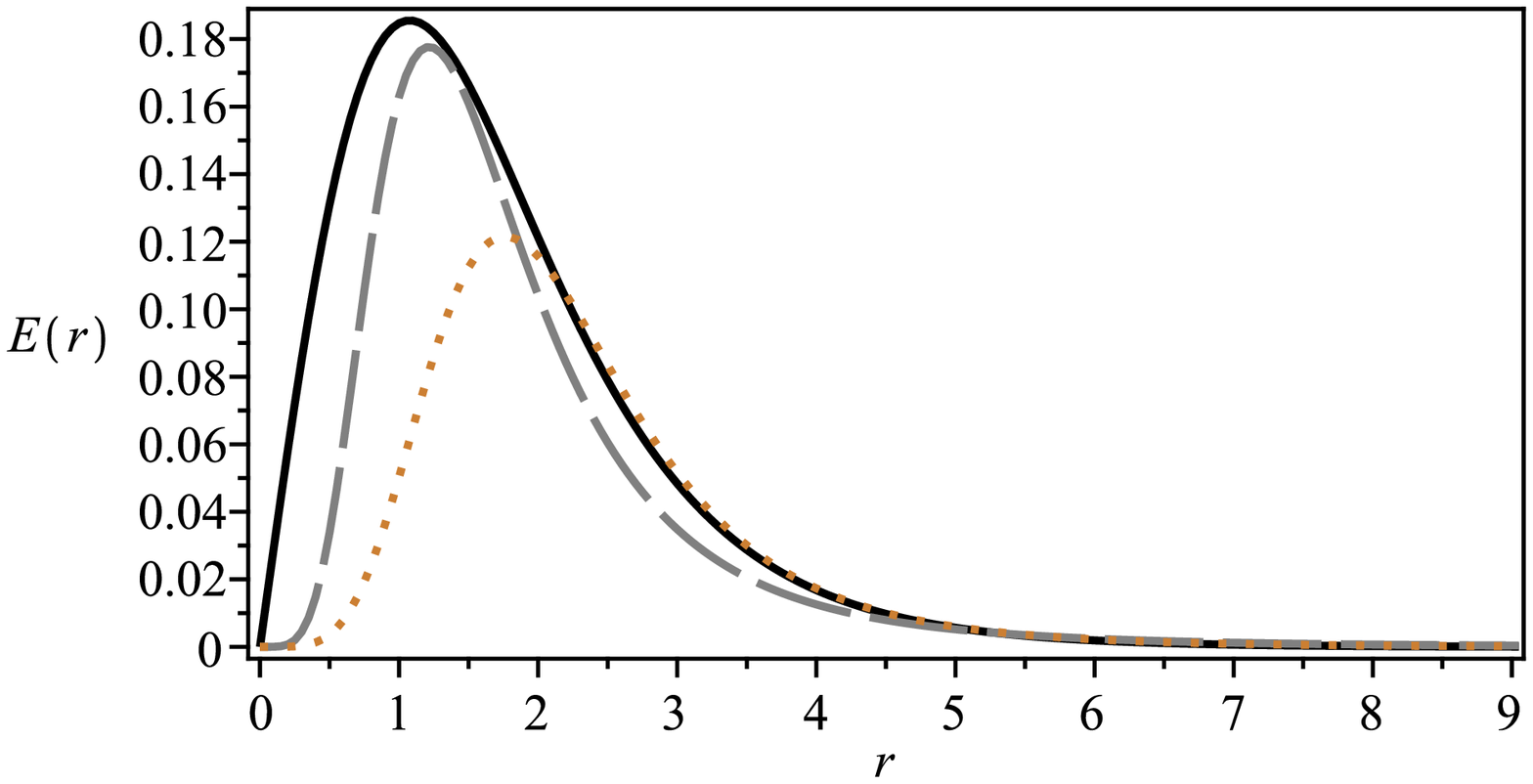} %
\includegraphics[width=8.4cm]{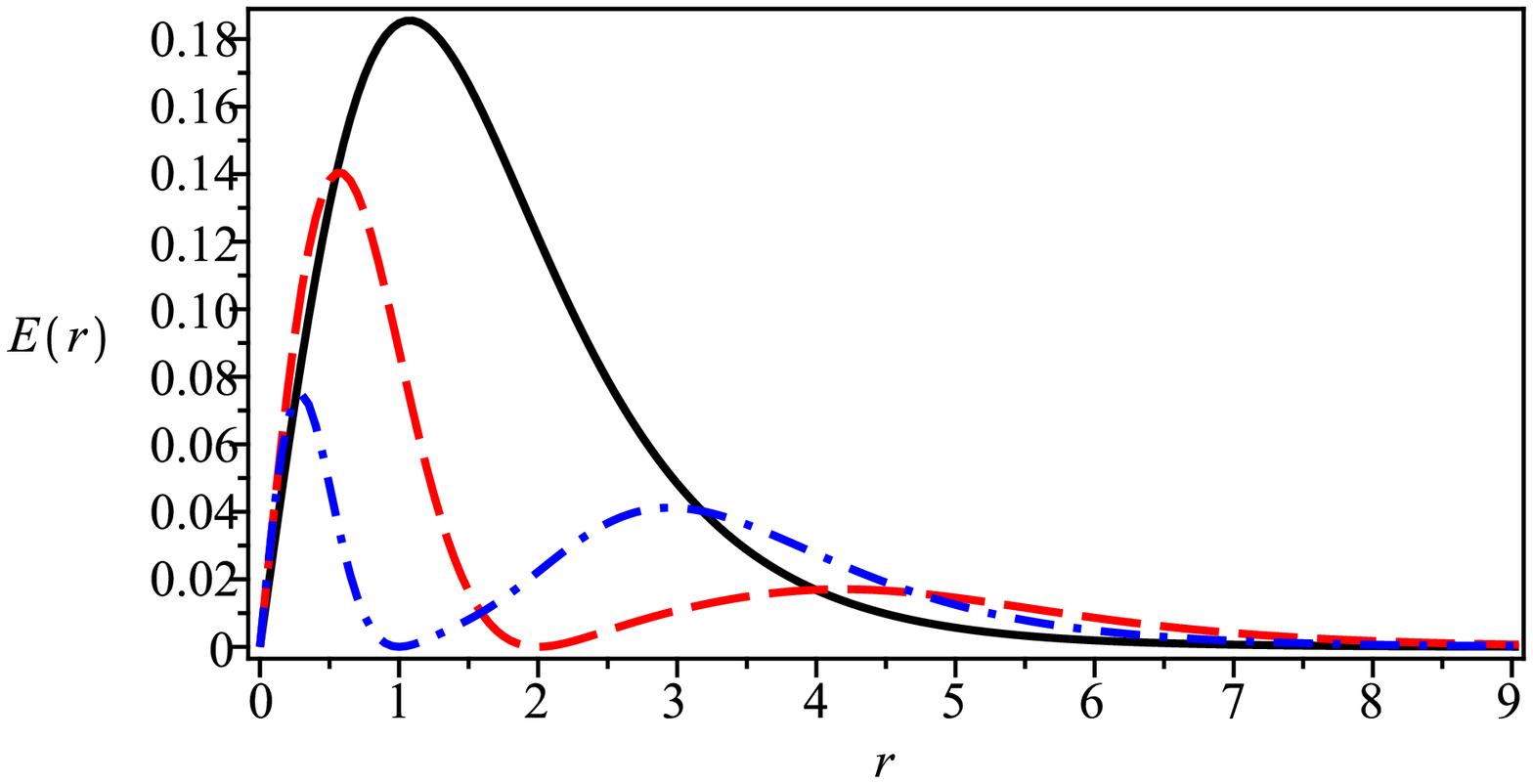}
\caption{Numerical solutions to the electric field $E(r)$. Conventions as in
the Fig. \ref{figg1}. The solutions for the 2nd. case vanish at $r=r_{0}$ (internal
structure) in order to avoid the second term in the right-hand side of the
Eq. (\protect\ref{ed}) to be singular.}
\end{figure}

We now proceed with the numerical study of the equations (\ref{b7}), (\ref%
{b8}) and (\ref{b9}) according the boundary values (\ref{b1}), (\ref{b2})
and (\ref{b3}). Here, we again choose $e=v=\kappa =n=1$, from which we solve
the equations for $r_{0}=1$ and $r_{0}=2$, see the dash-dotted blue and the
dashed red lines, respectively. Furthermore, with the aim to compare the new
results with the previous ones, we plot them in the same figures 1-6.

The profiles for the functions $g(r)$ and $a(r)$ naturally converge to the
boundary values (\ref{b1}) and (\ref{b2}), see the figures 1 and 2. The
novelty here is that the solutions for $a(r)$ now have a \textit{plateau} at
the point $r=r_{0}$. Here, it is worthwhile to note that the plateau at $%
r=r_{0}$ also occurs in the solutions for the electric potential $A_{0}(r)$,
see Figure 3. As we explain below, these plateaus give rise to the emergence
of the internal structure in the profiles for the magnetic and electric
fields.

In the figures 4 and 5, one finds the numerical profiles for the magnetic
and electric fields, respectively. In this case, we highlight the existence
of internal structures for intermediate values of the radial coordinate $r$
(in particular, the corresponding fields vanish at $r=r_{0}$). These
structures can be understood as being caused by the plateaus in the
solutions for $a(r)$ and $A_{0}(r)$ (remember that $B(r)\propto a^{\prime
}(r)$ and $E(r)\propto A_{0}^{\prime }(r)$). On the other hand, as we have
also explained previously, the existence of these structures avoid the first
two terms in the right-hand side of (\ref{ed}) to be singular and guarantees
that the energy of the respective vortices converges to the value in (\ref%
{te1}).
\begin{figure}[t]
\includegraphics[width=8.4cm]{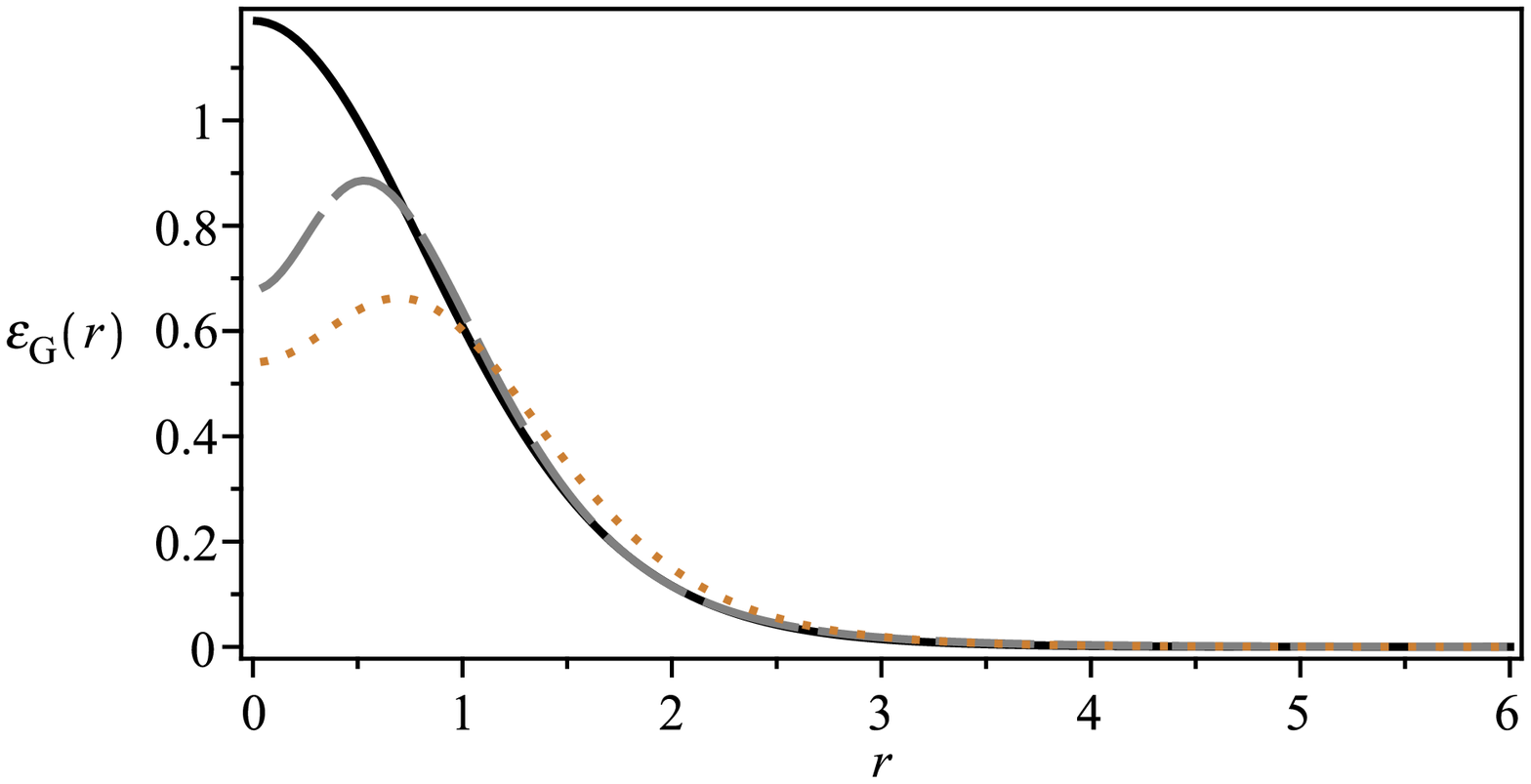} %
\includegraphics[width=8.4cm]{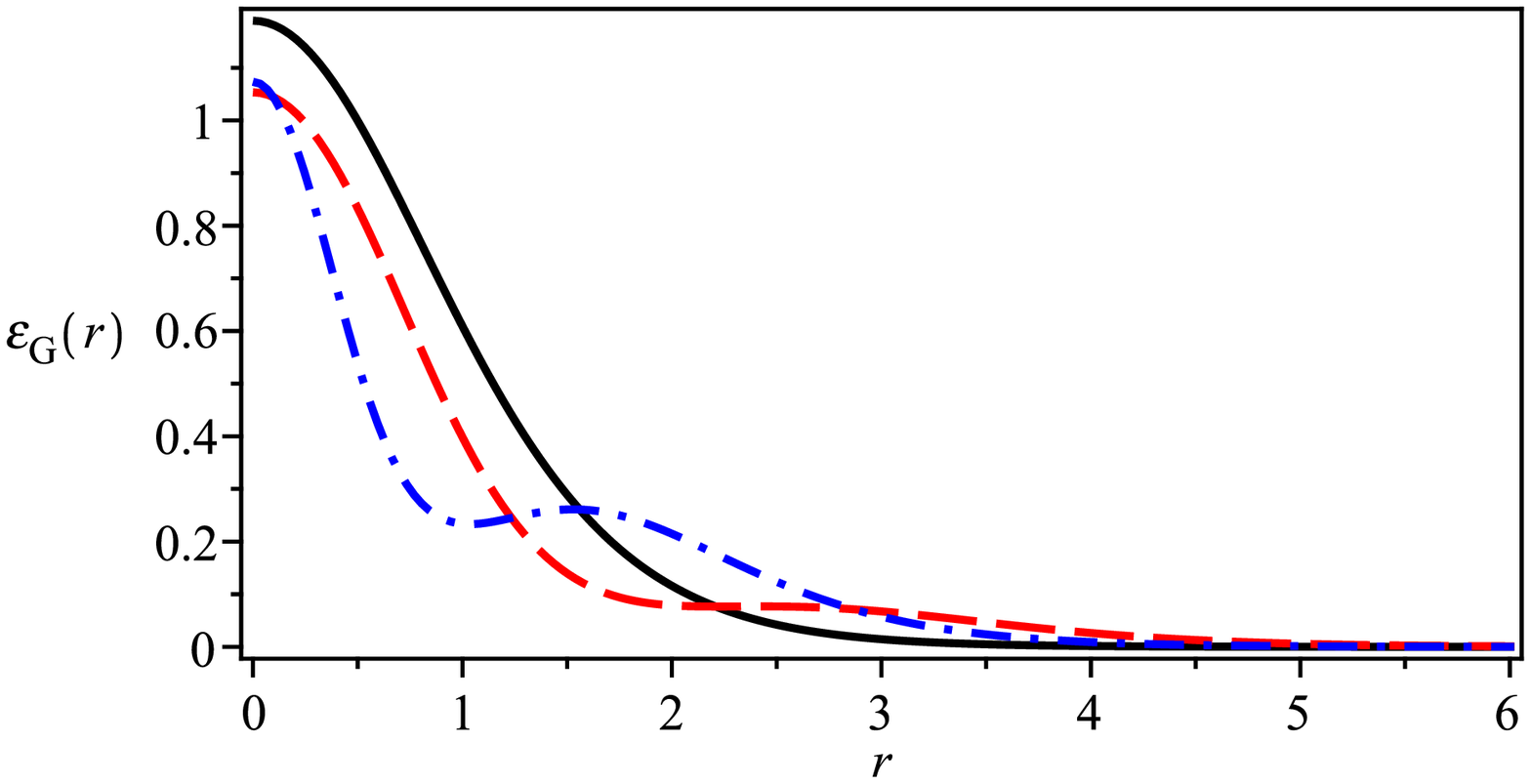}
\caption{Numerical solutions to the energy distribution $\protect\varepsilon %
_{G}$ related to the BPS configurations. Conventions as in the Fig. \ref{figg1}. Note
how the source field changes the shape of the resulting profiles.}
\end{figure}

The numerical results for the energy density $\varepsilon_{G}$ are shown in
the Fig. 6, from which we see how the source field again controls the shape
of the corresponding solutions, which eventually mimic the standard's shape,
see the solution for $r_{0}=2$ (i.e. the dashed red one).

We end this Section by verifying how the dielectric medium changes the way
the fields of the model behave near the boundaries. In this sense, we solve
the equations (\ref{b7}), (\ref{b8}) and (\ref{b9}) around the boundary
values (\ref{b1}), (\ref{b2}) and (\ref{b3}), from which we get that the
approximate solutions near the origin read (here, we have chosen $n=N>0$),%
\begin{equation}
g(r)\approx g_{N}r^{N}-\frac{eB_{0}}{4}g_{N}r^{N+2}\text{,}
\end{equation}%
\begin{equation}
a(r)\approx N-\frac{eB_{0}}{2}r^{2}\text{,}
\end{equation}%
\begin{equation}
A_{0}(r)\approx w_{0}-\frac{\kappa B_{0}}{4}r^{2}\text{,}
\end{equation}%
where $g_{N}>0$ and $w_{0}$ are integration constants, and $B_{0}$ is the
same parameter already defined in the Eq. (\ref{B00}). In the meantime, the
asymptotic solutions can be verified to be given by%
\begin{equation}
g(r)\approx 1-C_{\infty }\frac{e^{-Mr}}{\sqrt{r}}\text{,}
\end{equation}%
\begin{equation}
a(r)\approx MC_{\infty }\sqrt{r}e^{-Mr}\text{,}
\end{equation}%
\begin{equation}
A_{0}(r)\approx \frac{MC_{\infty }}{e}\frac{e^{-Mr}}{\sqrt{r}}\text{,}
\end{equation}%
which therefore recover the typical behavior (i.e., the exponential decay).
In this case, we have identified%
\begin{equation}
M=\frac{1}{2}\sqrt{8e^{2}v^{2}+\kappa ^{2}}-\frac{|\kappa |}{2}\text{,}
\end{equation}%
as the mass of the corresponding bosons. The important observation here is
that the dielectric medium (\ref{tg2}) does not change the manner the
fundamental fields approach the boundaries (remember that the medium equals
the unity at the boundaries, i.e. $G(r=0)=1$ and $G(r\rightarrow \infty
)\rightarrow 1$). In this sense, the basic fields are expected to mimic the
behavior of the ones obtained from the self-dual Maxwell-Chern-Simons-Higgs
model in the \textit{absence} of the dielectric medium.

%%%%%%%%%%%%%%%%%%%%%%%%

\section{The Maxwell-Chern-Simons-$CP(2)$ case \label{2 copy(1)}}

We now go further in our investigation by focusing our attention on the
first-order vortices with internal structure inherent to an extended
Maxwell-Chern-Simons-$CP(2)$ theory. Here, it is important to say that the
observations introduced during the construction of the self-dual MCSH theory
apply again in order to generate a MCS-$CP(2)$ self-dual scenario \cite%
{neyver}, including the one related to the formation of the internal
structures.

We then begin by considering the Lagrange density which describes the
extended MCS-$CP(2)$ model, i.e.%
\begin{eqnarray}
\mathcal{L} &=&-\frac{G(\chi )}{4}F_{\mu \upsilon }F^{\mu \upsilon }-\frac{%
\kappa }{4}\epsilon ^{\beta \mu \upsilon }A_{\beta }F_{\mu \upsilon }  \notag
\\
&&+\overline{\left( P_{ab}D_{\mu }\phi _{b}\right) }\left( P_{ac}D^{\mu
}\phi _{c}\right) +\frac{G(\chi )}{2}\partial _{\mu }\Psi \partial ^{\mu
}\Psi  \notag \\
&&+\frac{1}{2}\partial _{\mu }\chi \partial ^{\mu }\chi -U\left( \phi
_{3},\Psi ,\chi \right) \text{,}  \label{m1}
\end{eqnarray}%
where the Latin indexes count the components of the $CP(2)$ sector, $\phi
_{a}$ represents the $CP(2)$ field which is constrained to satisfy the
condition $\phi _{a}\overline{\phi }_{a}=h$ (the constant $h$ represents the
norm of the field), $P_{ab}=\delta _{ab}-h^{-1}\phi _{a}\overline{\phi }_{b}$
stands for a projector in the internal space and $D_{\mu }\phi _{a}$ is the
covariant derivative of the model, i.e.%
\begin{equation}
D_{\mu }\phi _{a}=\partial _{\mu }\phi _{a}-igA_{\mu }Q_{ab}\phi _{b}\text{,}
\end{equation}%
where $g$ is a coupling constant and $Q_{ab}$ stands for a charge matrix
that, for our purpose, we choose as%
\begin{equation}
Q_{ab}=\frac{1}{2}\text{diag}\left( 1,-1,0\right) \text{.}
\end{equation}%
The remaining definitions and conventions are the same as in the previous
Section.

The potential reads as%
\begin{eqnarray}
U\left( \phi _{3},\Psi ,\chi \right) &=&\frac{1}{2G}\left( h^{1/2}g\phi
_{3}+\kappa \Psi \right) ^{2}  \notag \\[0.2cm]
&&+g^{2}\Psi ^{2}\overline{\left( P_{ab}Q_{bm}\phi _{m}\right) }%
P_{ac}Q_{cn}\phi _{n}  \notag \\
&&+\frac{1}{2r^{2}}\left( \frac{dW}{d\chi }\right) ^{2}\text{,}  \label{mp}
\end{eqnarray}%
where the first two terms in the right-hand side (with $G=1$) represent the
self-dual potential inherent to the canonical MCS-$CP(2)$ theory \cite%
{neyver}. The dependence of the potential on the third component of the $%
CP(2)$ field presupposes the spontaneous breaking of $SU(3)$-symmetry
inherent to the model. On the other hand, the presence of the Chern-Simons
term justifies the obtainment of solutions with internal structure carrying
on both electric and magnetic fields.

We consider time-independent configurations described by the map%
\begin{equation}
A_{i}=-\frac{1}{gr^{2}}\varepsilon _{ij}x_{j}A(r)\text{,}  \label{m2}
\end{equation}%
\begin{equation}
\left(
\begin{array}{c}
\phi _{1} \\
\phi _{2} \\
\phi _{3}%
\end{array}%
\right) =h^{\frac{1}{2}}\left(
\begin{array}{c}
e^{im\theta }\sin \alpha \cos \beta \\
e^{-im\theta }\sin \alpha \sin \beta \\
\cos \alpha%
\end{array}%
\right) \text{,}  \label{m3}
\end{equation}%
where $\alpha =\alpha (r)$ and $\beta =\beta (r)$ are a priori functions of
the radial coordinate $r$ only. Here, the vorticity of the final solutions
is determined by the winding number $m=\pm 1,\,\pm 2,\,\pm 3...$, while the
above map again leads us to $A_{0}=A_{0}(r)$, $\Psi =\Psi (r)$ and $\chi
=\chi (r)$.

It is useful to emphasize that the presence of the source field $\chi $ does
not change the equation of motion for $\beta (r)$ which still reads as in
the Eq. (17) of the Ref. \cite{neyver}. Thus, one again gets that the
simplest solutions for $\beta (r)$ are%
\begin{equation}
\beta _{1}=\frac{\pi }{4}+\frac{n\pi }{2}\text{ \ and \ }\beta _{2}=\frac{%
n\pi }{2}\text{,}
\end{equation}%
with $n\in \mathbb{Z}$. In this section, we work with the first choice only,
i.e. $\beta (r)=\beta _{1}$.

Furthermore, the other field profiles $\alpha (r)$ and $A(r)$ are supposed
to satisfy the same boundary conditions as they do in the absence of the
dielectric medium, i.e.%
\begin{equation}
\alpha \left( r=0\right) =0\text{ \ and \ }A\left( r=0\right) =0\text{,}
\label{m4}
\end{equation}%
\begin{equation}
\alpha \left( r\rightarrow \infty \right) \rightarrow \frac{\pi }{2}\text{ \
and \ }A\left( r\rightarrow \infty \right) \rightarrow 2m\text{,}  \label{m5}
\end{equation}%
which are known to lead to well-behaved solutions with finite energy.

We proceed with the minimization of the total energy again via the
implementation of the BPS algorithm. In order to attain this goal, we first
write the expression for the energy density related to the model (\ref{m1}),
which reads%
\begin{eqnarray}
\varepsilon &=&\frac{G}{2}B^{2}+\frac{G}{2}\left( \frac{dA_{0}}{dr}\right)
^{2}  \notag \\
&&+h\left( \left( \frac{d\alpha }{dr}\right) ^{2}+\frac{\left( 2m-A\right)
^{2}}{4r^{2}}\sin ^{2}\alpha \right)  \notag \\
&&+\frac{g^{2}h}{4}\left( A_{0}\right) ^{2}\sin ^{2}\alpha +\frac{G}{2}%
\left( \frac{d\Psi }{dr}\right) ^{2}  \notag \\
&&+\frac{1}{2}\left( \frac{d\chi }{dr}\right) ^{2}+\frac{1}{2G}\left( hg\cos
\alpha +\kappa \Psi \right) ^{2}  \notag \\
&&+\frac{h}{4}g^{2}\Psi ^{2}\sin ^{2}\alpha +\frac{1}{2r^{2}}\left( \frac{dW%
}{d\chi }\right) ^{2}\text{,}  \label{ed1}
\end{eqnarray}%
where we have used the Eq. (\ref{mp}) for the potential $U=U(\phi _{3},\Psi
,\chi )$ and the radially symmetric map (\ref{m2}) and (\ref{m3}), with $%
\beta (r)=\beta _{1}$. In the sequence, after some algebraic work, the above
expression can be written in the form%
\begin{eqnarray}
\varepsilon &=&\varepsilon _{BPS}+\frac{1}{2G}\left( \frac{{}}{{}}GB\mp
\left( hg\cos \alpha +\kappa \Psi \right) \right) ^{2}  \notag \\
&&+h\left[ \frac{d\alpha }{dr}\mp \frac{\left( 2m-A\right) }{2r}\sin \alpha %
\right] ^{2}+\frac{1}{2}\left( \frac{d\chi }{dr}\mp \frac{1}{r}\frac{dW}{%
d\chi }\right) ^{2}\   \notag \\
&&+\frac{1}{2}G\left( \frac{dA_{0}}{dr}\pm \frac{d\Psi }{dr}\right) ^{2}+%
\frac{g^{2}h}{4}\left( A_{0}\pm \Psi \right) ^{2}\sin ^{2}\alpha \text{,}
\end{eqnarray}%
in which we have again introduced the term $\varepsilon _{BPS}$ which is now
given by%
\begin{equation}
\varepsilon _{BPS}=\mp \frac{1}{r}\frac{d}{dr}\left[ h\left( 2m-A\right)
\cos \alpha \right] \pm \frac{1}{r}\frac{dW}{dr}\text{,}  \label{ed1c}
\end{equation}%
whose integration over the plane, using the boundary values (\ref{m4}) and (%
\ref{m5}), provides the quantity that stands for the BPS energy of the model
(\ref{m1}), i.e.%
\begin{equation}
E_{BPS}=2\pi \int_{0}^{\infty }r\varepsilon _{BPS}dr=\pm 2\pi \left(
2mh+\Delta W\right) \text{,}  \label{te2}
\end{equation}%
where $\Delta W=W\left( r\rightarrow \infty \right) -W(r=0)$, while the
upper (lower) sign holds for $m>0$ and $\Delta W>0$ ($m<0$ and $\Delta W<0$).

The construction above reveals that the\ total\ energy\ satisfies%
\begin{eqnarray}
&&\left. \frac{E}{2\pi }=\int_{0}^{\infty }\frac{1}{2G}\left( \frac{{}}{{}}%
GB\mp \left( hg\cos \alpha +\kappa \Psi \right) \right) ^{2}rdr\right.
\notag \\
&&\left. +\int_{0}^{\infty }\left[ h\left( \frac{d\alpha }{dr}\mp \frac{%
\left( 2m-A\right) }{2r}\sin \alpha \right) ^{2}\right. \right.  \notag \\
&&\left. \text{ \ \ \ \ \ \ \ \ \ \ \ \ \ \ \ \ \ \ \ \ \ \ \ \ \ \ \ \ \ }%
\left. +\frac{1}{2}\left( \frac{d\chi }{dr}\mp \frac{1}{r}\frac{dW}{d\chi }%
\right) ^{2}\right] rdr\right.  \notag \\
&&\left. +\int_{0}^{\infty }\left[ \frac{G}{2}\left( \frac{dA_{0}}{dr}\pm
\frac{d\Psi }{dr}\right) ^{2}+\frac{g^{2}h}{4}\left( A_{0}\pm \Psi \right)
^{2}\sin ^{2}\alpha \right] rdr\right.  \notag \\
&&\left. +\frac{E_{BPS}}{2\pi }\geq 2h\left\vert m\right\vert +\left\vert
\Delta W\right\vert \right.
\end{eqnarray}%
and that it is therefore bounded from below. In this sense, the fields which
saturate the bound are necessarily the ones which satisfy the following
system of equations%
\begin{equation}
GB=\pm hg\cos \alpha \pm \kappa \Psi \text{,}  \label{bpsx1}
\end{equation}%
\begin{equation}
\frac{d\alpha }{dr}=\pm \frac{\left( 2m-A\right) }{2r}\sin \alpha \text{,}
\label{bpsx2}
\end{equation}%
\begin{equation}
\frac{d\chi }{dr}=\pm \frac{1}{r}\frac{dW}{d\chi }\text{,}  \label{qiqi}
\end{equation}%
\begin{equation}
\frac{dA_{0}}{dr}=\mp \frac{d\Psi }{dr}\text{ \ and \ }A_{0}=\mp \Psi \text{,%
}
\end{equation}%
which provides the first-order equations inherent to the effective theory.
In other words, the fields which solve these differential equations
originate BPS vortices whose total energy is given by the Eq. (\ref{te2}),
i.e. the corresponding Bogomol'nyi bound.
\begin{figure}[t]
\includegraphics[width=8.4cm]{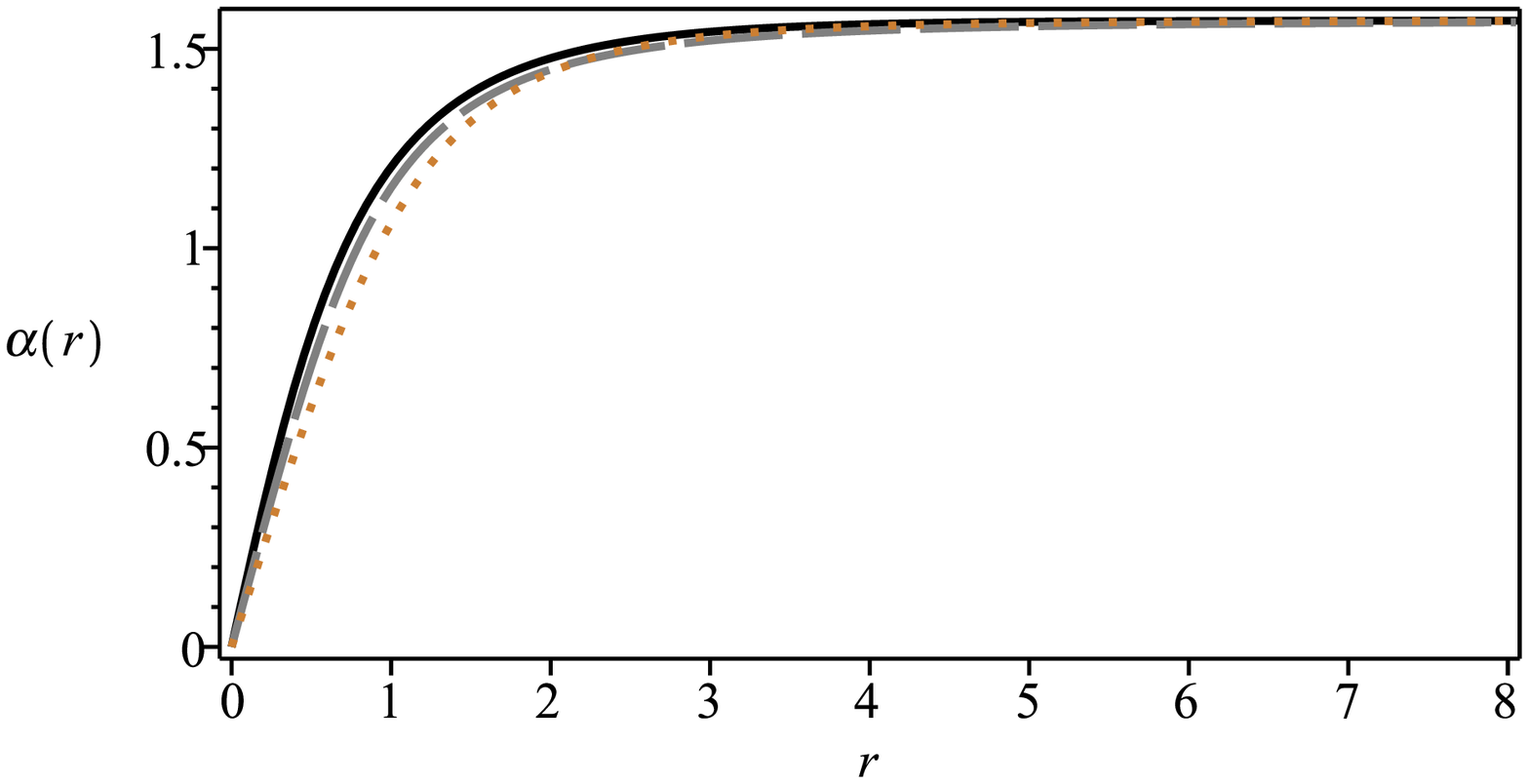} %
\includegraphics[width=8.4cm]{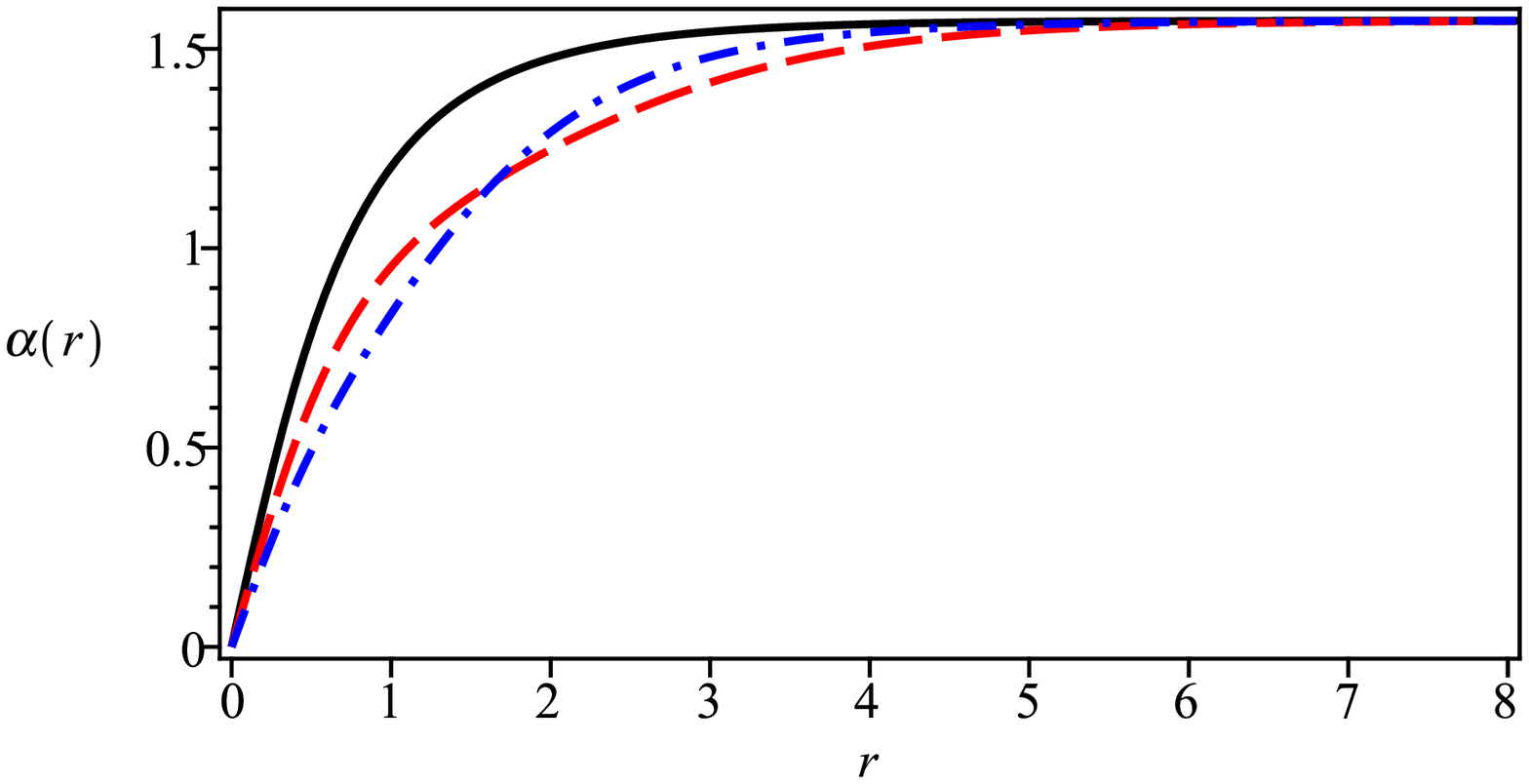}
\caption{Numerical solutions to the scalar profile function $\protect\alpha %
(r)$ coming from (\protect\ref{xx1}), (\protect\ref{xx2}) and (\protect\ref%
{xx3}) (top, long-dashed gray line for $r_{0}=1$ and dotted orange line for $%
r_{0}=2$)\ and (\protect\ref{y1}), (\protect\ref{y2}) and (\protect\ref{y3})
(bottom, dash-dotted blue line for $r_{0}=1$ and dashed red line for $%
r_{0}=2 $). In both cases, the equations were solved according the boundary
conditions (\protect\ref{m4}), (\protect\ref{m5}) and (\protect\ref{b3}). We
have used $h=\protect\kappa =1$,\ $g=2$ and $m=1$. The solid black line
represents the canonical solution (no source field, with $G=1$).} \label{figg7}
\end{figure}

In the same way as in the previous case, the last first-order equation, $%
\Psi =\mp A_{0}$, solves the penultimate one, $\Psi ^{\prime }=\mp
A_{0}^{\prime }$, identically. In addition, the electric potential must be
again obtained by solving the time-independent Gauss law,%
\begin{equation}
\frac{1}{r}\frac{d}{dr}\left( rG\frac{dA_{0}}{dr}\right) +\kappa B=\frac{%
g^{2}h}{2}A_{0}\sin ^{2}\alpha \text{,}
\end{equation}%
with $A_{0}(r)$ satisfying the same boundary conditions given in the Eq. (%
\ref{b3}).

Furthermore, it is possible to express the BPS energy density (\ref{ed1c})
in the same way as in (\ref{ed1xcp}), with $\varepsilon _{G}$ representing
the contribution which comes from the $CP(2)$ soliton with internal
structure, i.e.%
\begin{eqnarray}
\varepsilon _{G} &=&G\left[ B^{2}+\left( \frac{dA_{0}}{dr}\right) ^{2}\right]
\notag \\
&&\text{ \ \ \ \ }+\frac{h}{2}\left[ g^{2}\left( A_{0}\right) ^{2}+\frac{%
(2m-A)^{2}}{r^{2}}\right] \sin ^{2}\alpha \text{,}  \label{edGcp}
\end{eqnarray}%
and $\varepsilon _{\chi }$ standing for the energy density of the source
field $\chi $ given again by the Eq. (\ref{edX}).

\subsection{Some MCS-$CP(2)$ scenarios with internal structure}

We proceed in the same way as in the Sec. \ref{2} in order to introduce some
first-order scenarios. In this sense, for comparison purposes, we choose the
superpotential $W(\chi )$ again as it appears in the Eq. (\ref{w}) of the
previous Sec. \ref{MCSHsc}, i.e.%
\begin{equation}
W(\chi )=\chi -\frac{1}{3}\chi ^{3}\text{,}
\end{equation}%
which, in view of the Eq. (\ref{qiqi}), provides the solution for the source
field as given in the Eq. (\ref{ssf}), i.e.%
\begin{equation}
\chi (r)=\pm \frac{r^{2}-r_{0}^{2}}{r^{2}+r_{0}^{2}}\text{.}  \label{ssf2}
\end{equation}

\subsubsection{The first case}

Now, we specify the dielectric function $G(\chi )$. Here, in order to to
compare the corresponding effects with the ones observed previously in the
MCSH case, we choose this function as in the Eq. (\ref{tg1}) or,
alternatively, as in the Eq. (\ref{tg11}),%
\begin{equation}
G(\chi )=\frac{1}{1-\chi ^{2}}=\frac{\left( r^{2}+r_{0}^{2}\right) ^{2}}{%
4r^{2}r_{0}^{2}}\text{,}  \label{gg}
\end{equation}%
whose peculiarities were already discussed in the previous Section \ref%
{fcmcsh1}.

In view of (\ref{gg}), the first-order equations (\ref{bpsx1}) and (\ref%
{bpsx2}) reduce to%
\begin{equation}
\frac{1}{r}\frac{dA}{dr}=\frac{4r^{2}r_{0}^{2}}{\left(
r^{2}+r_{0}^{2}\right) ^{2}}\left( \pm hg^{2}\cos \alpha -g\kappa
A_{0}\right) \text{,}  \label{xx1}
\end{equation}%
\begin{equation}
\frac{d\alpha }{dr}=\pm \frac{\left( 2m-A\right) }{2r}\sin \alpha \text{,}
\label{xx2}
\end{equation}%
while the Gauss law assumes the form%
\begin{equation}
\frac{1}{4r_{0}^{2}r}\frac{d}{dr}\left[ \frac{\left( r^{2}+r_{0}^{2}\right)
^{2}}{r}\frac{dA_{0}}{dr}\right] +\kappa B=\frac{g^{2}h}{2}A_{0}\sin
^{2}\alpha \text{,}  \label{xx3}
\end{equation}%
where $B$ represents the magnetic field given by%
\begin{equation}
B(r)=\frac{1}{gr}\frac{dA}{dr}\text{.}  \label{Bcp2}
\end{equation}

We perform the numerical analysis again by using a finite-difference scheme
and according the conditions (\ref{m4}), (\ref{m5}) and (\ref{b3}). Here, we
choose $h=\kappa =1$, $g=2$, and $m=1$ (upper signs in the first-order
expressions). We then solve the BPS equations (\ref{xx1}), (\ref{xx2}) and (%
\ref{xx3}) for $r_{0}=1$ (long-dashed gray line) and $r_{0}=2$ (dotted
orange line), from which we depict the numerical profile for the fields $%
\alpha (r)$ and $A(r)$, the electric potential $A_{0}(r)$, the magnetic $%
B(r) $ and electric $E(r)$ fields, and the energy distribution $\varepsilon
_{G}$, see the figures 7-12 (the solid black line stands for the solution
with no source field, i.e. $G=1$).
\begin{figure}[t]
\includegraphics[width=8.4cm]{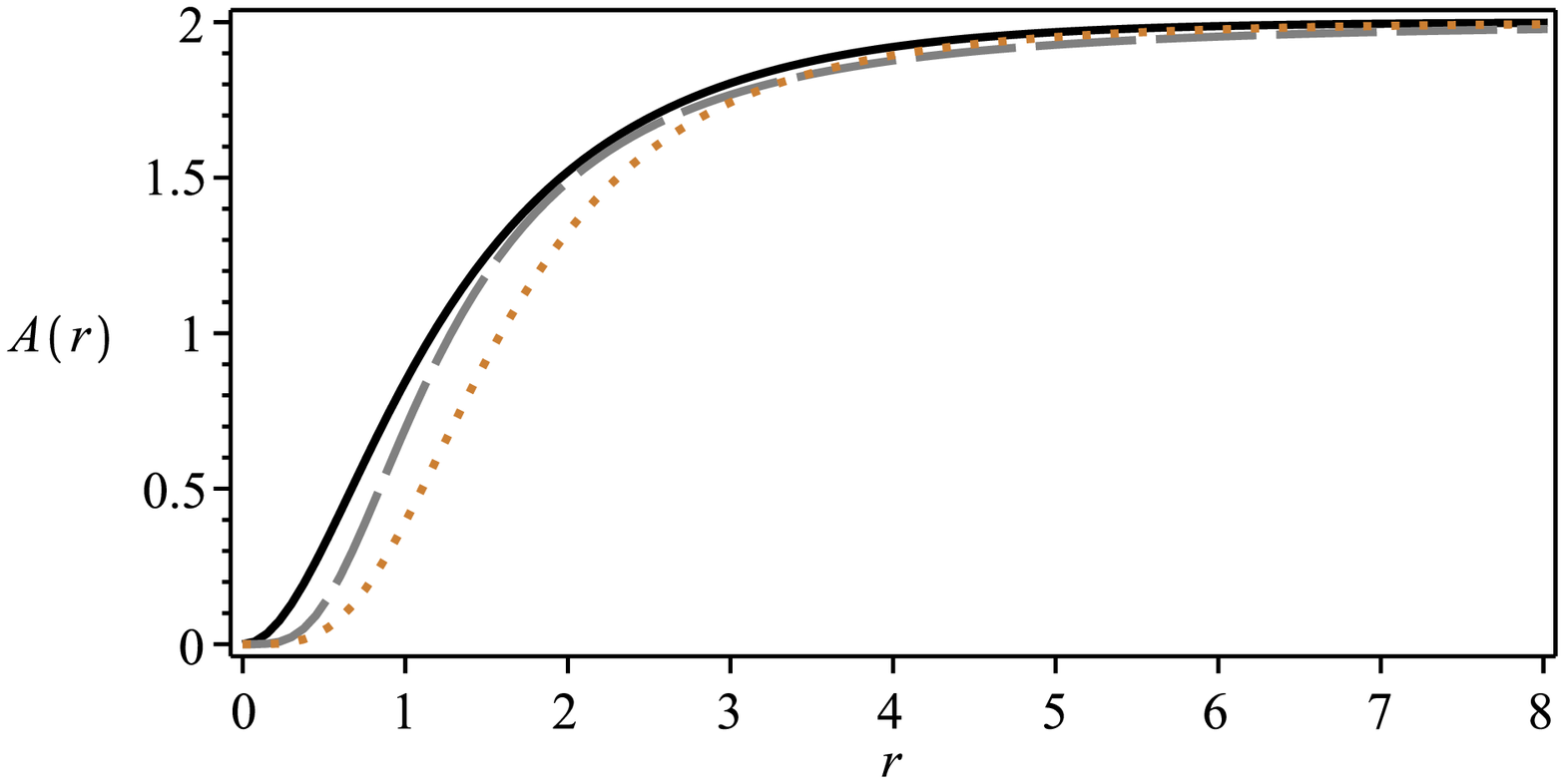} %
\includegraphics[width=8.4cm]{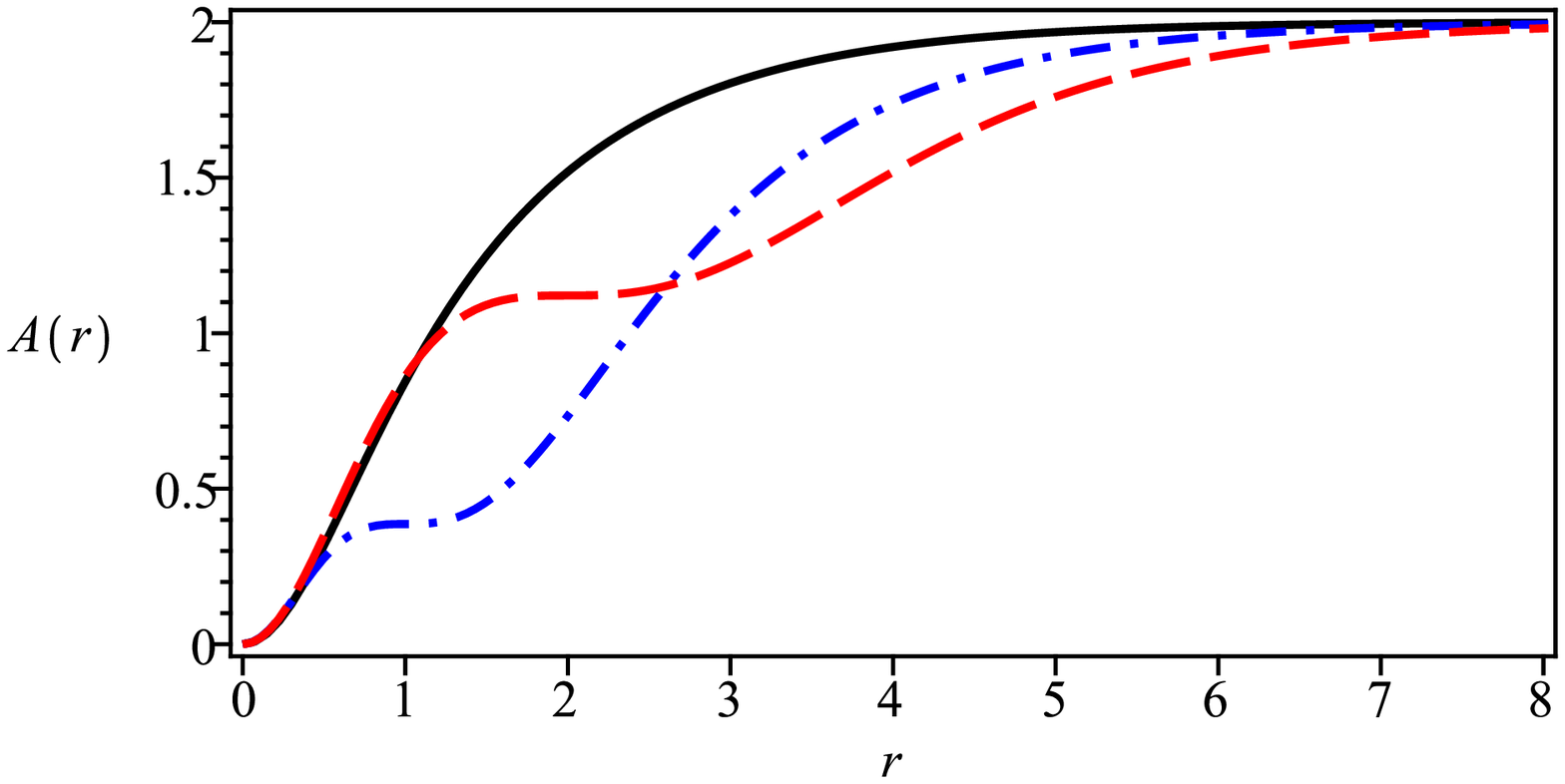}
\caption{Numerical solutions to the gauge profile function $A(r)$.
Conventions as in the Fig. \ref{figg7}. The solutions for the 1st. case mimic the
usual ones around $r=r_{0}$, while the profiles for the 2nd. case present a
plateau at the same point.}
\end{figure}

In this sense, the field profiles $\alpha (r)$ and $A(r)$ are shown in the
figures 7 and 8, respectively. In view of them, it's possible to note that
the presence of the source field causes slight variations on the respective
core-size. Also here, the profiles converge to the boundary values (\ref{m4}%
) and (\ref{m5}) despite the presence of the dielectric medium itself.

The Figure 9 shows the solutions for the electric potential $A_{0}(r)$, from
which one notes that the behavior of $A_{0}(r=0)=w_{0}$ as a function of $%
r_{0}$ is the very similar to the one which we have already encountered in
the previous MCSH case (see the Fig. 3), i.e. as $r_{0}$ increases, $w_{0}$
decreases.
\begin{figure}[t]
\includegraphics[width=8.4cm]{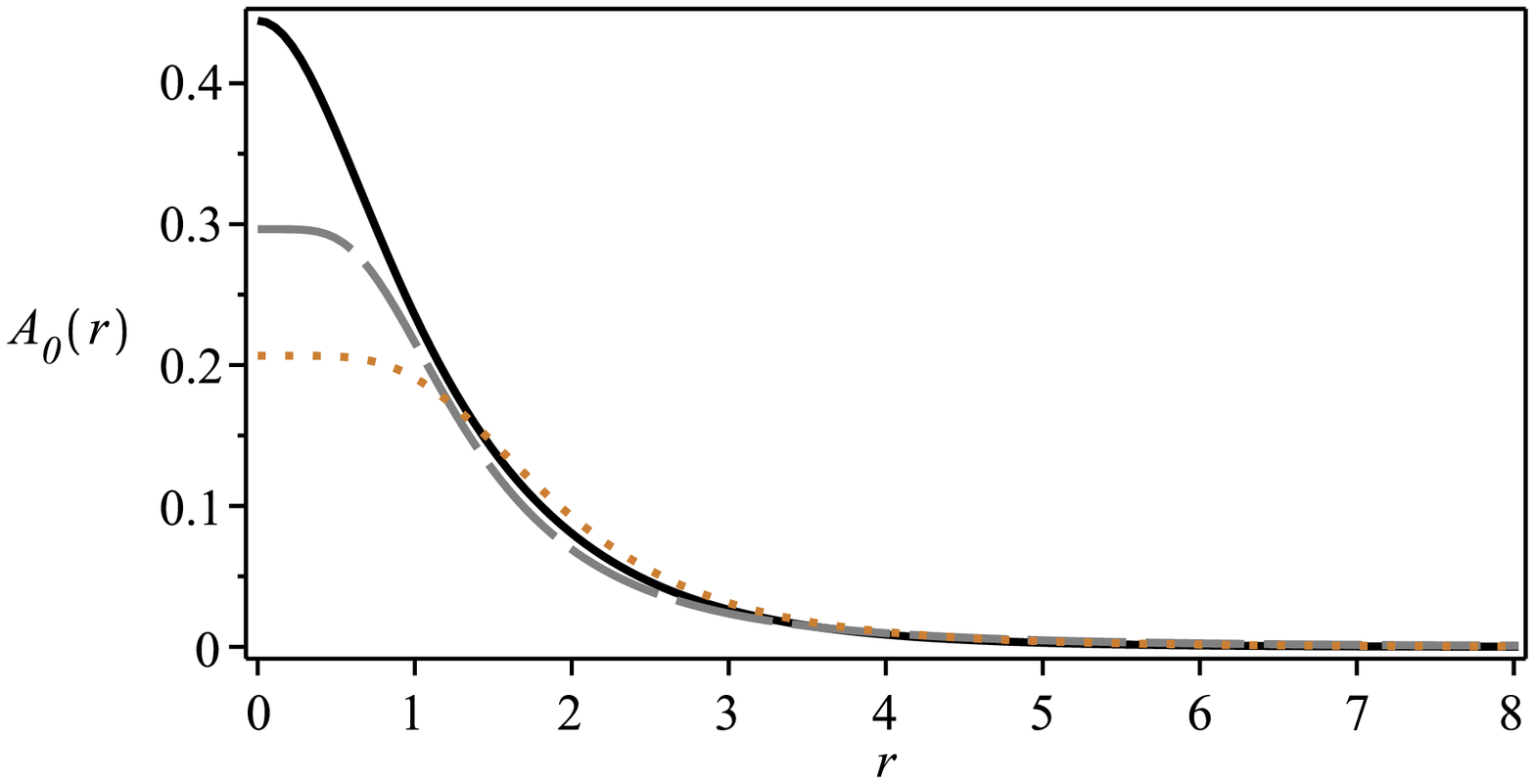} %
\includegraphics[width=8.4cm]{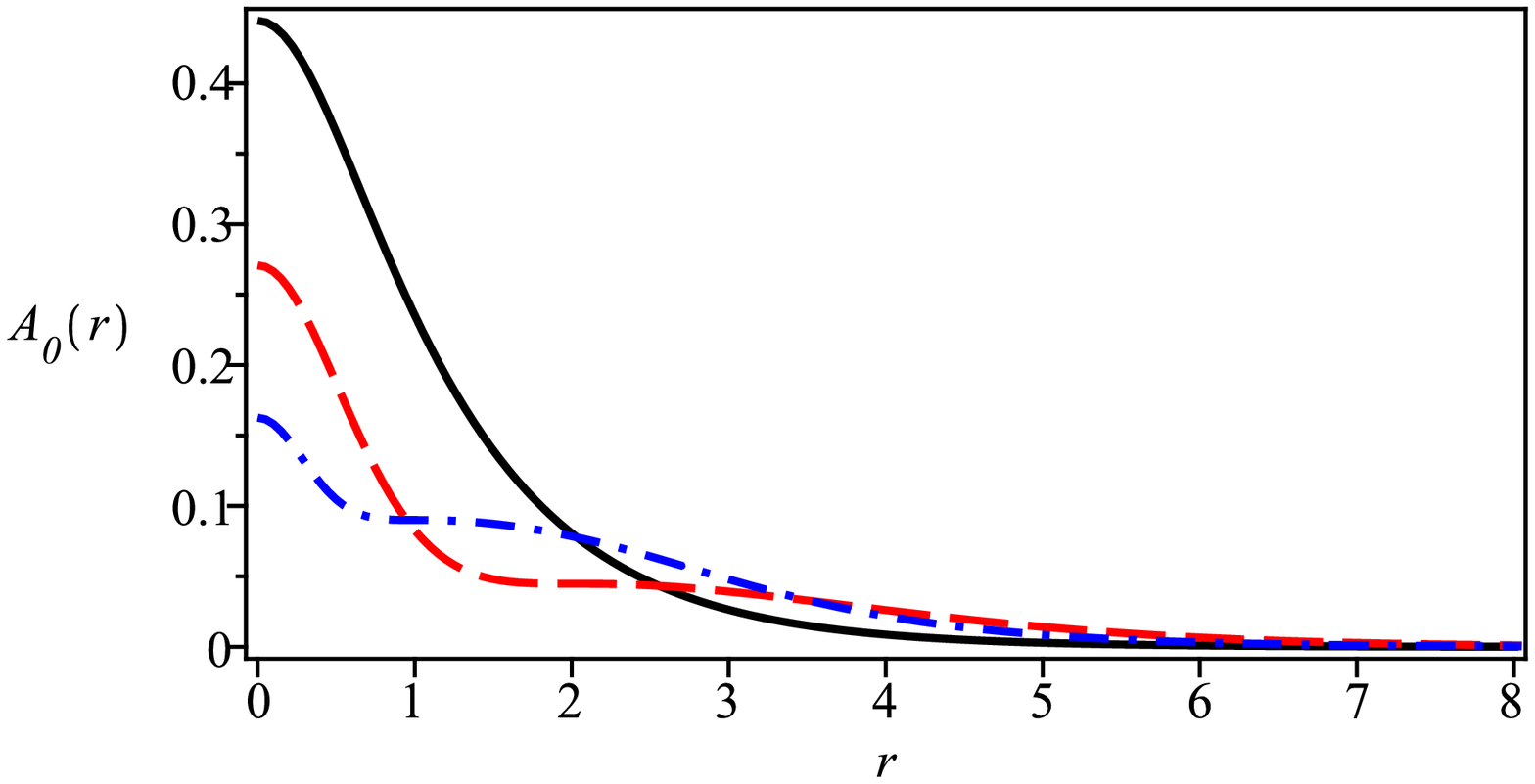}
\caption{Numerical solutions to the electric potential $A_{0}(r)$.
Conventions as in the \ref{figg7}. The solutions for the 1st. case mimic the
standard behavior around $r=r_{0}$. The results for the 2nd. case also have
a plateau at the same point.}
\end{figure}

The solutions for the magnetic and electric fields can be seen in figures 10
and 11, respectively. We see that the dielectric medium causes the very same
effect already observed in the MCSH case, i.e. the magnetic and electric
fields vanish at the boundaries to ensure that the energy of the resulting
BPS vortices converges to the bound given by the Eq. (\ref{te2}).

The Figure 12 presents the solutions for the energy density $\varepsilon
_{G} $ of the first-order vortices, via which one gets that again the
presence of the source field changes the shape of the resulting profiles.
Moreover, as it happens for the electric potential, as the values of $r_{0}$
increase, the values of $\varepsilon _{G}(r=0)$ decrease.

We finalize the study of this case by solving the first-order system (\ref%
{xx1}), (\ref{xx2}) and (\ref{xx3}) near to the boundary conditions (\ref{m4}%
), (\ref{m5}) and (\ref{b3}). So, around the boundary $r=0$, we solve for $%
m=1$ (the case analyzed numerically), which provides the following behavior
for the field profiles:%
\begin{equation}
\alpha (r)\approx \alpha _{1}r-\frac{\left( \alpha _{1}\right) ^{3}}{12}r^{3}%
\text{,}
\end{equation}%
\begin{equation}
A(r)\approx \frac{g\tilde{B}_{0}}{r_{0}^{2}}r^{4}-\left( \frac{gA_{1}}{%
3r_{0}^{2}}+\frac{4g\tilde{B}_{0}}{3r_{0}^{4}}\right) r^{6}\text{,}
\end{equation}%
\begin{equation}
A_{0}(r)\approx w_{0}+w_{2}r^{2}\text{,}
\end{equation}%
where $\alpha _{1}$ and $w_{2}$ stand for positive integration constants,
with $A_{1}=2w_{2}\kappa +gh\left( \alpha _{1}\right) ^{2}$ and
\begin{figure}[t]
\includegraphics[width=8.4cm]{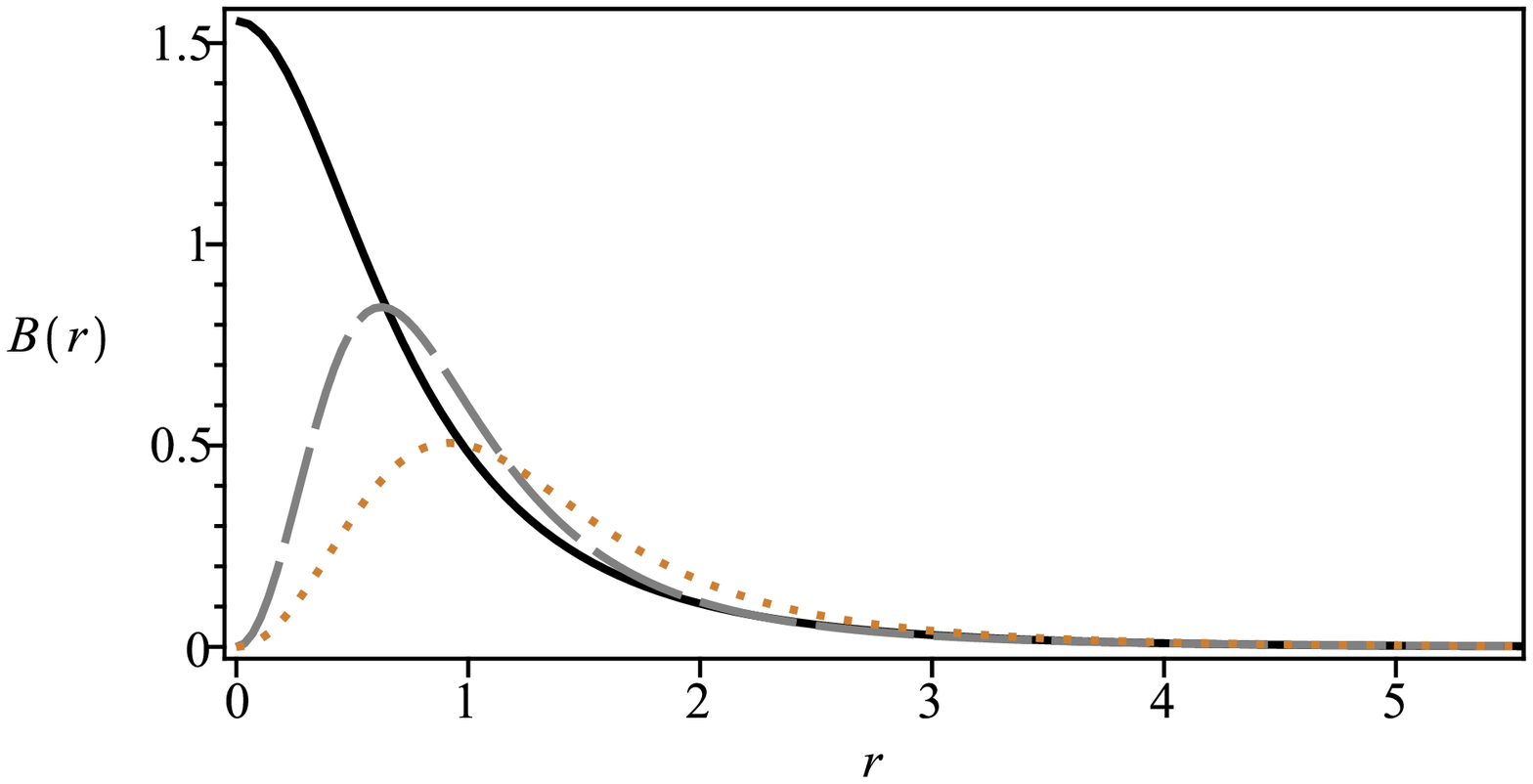} %
\includegraphics[width=8.4cm]{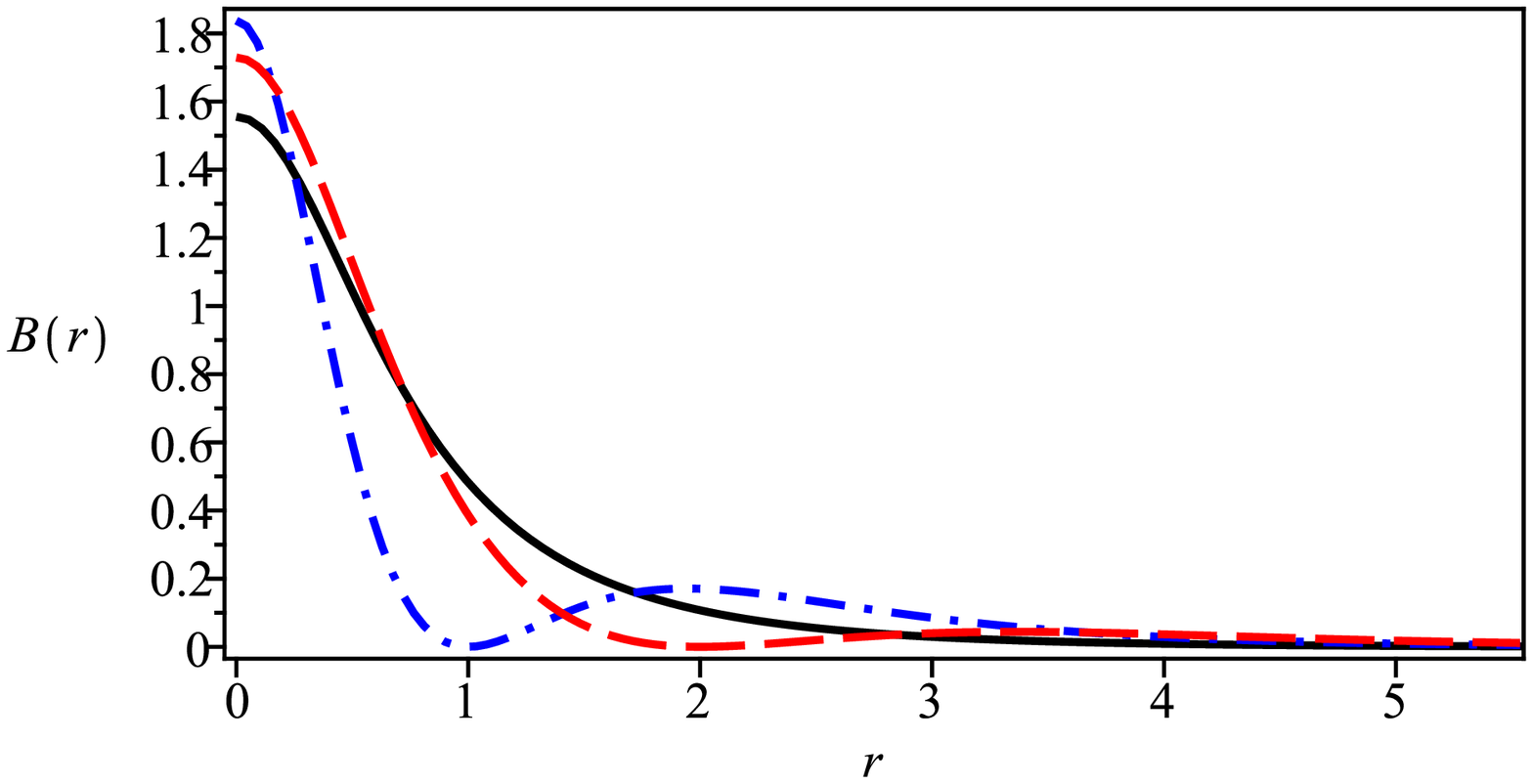}
\caption{Numerical solutions to the magnetic field $B(r)$. Conventions as in
the \ref{figg7}. The solutions for the 2nd. case vanish at $r=r_{0}$ in order to
avoid the first term in the right-hand side of the Eq. (\protect\ref{ed1})
to be singular.}
\end{figure}
\begin{equation}
\tilde{B}_{0}=hg-w_{0}\kappa \text{,}  \label{BBcp2}
\end{equation}%
which stands for the value of the magnetic field at $r=0$ for the MCS-$CP(2)$
model in the absence of the dielectric medium. On the other hand, in the
limit $r\rightarrow \infty $, we obtain%
\begin{equation}
\alpha (r)\approx \frac{\pi }{2}-C_{\infty }r^{-\Lambda }\text{,}
\end{equation}%
\begin{equation}
A(r)\approx 2m-2\Lambda C_{\infty }r^{-\Lambda }\text{,}
\end{equation}%
\begin{equation}
A_{0}(r)\approx -\frac{\Lambda C_{\infty }}{g\kappa \left( r_{0}\right) ^{2}}%
r^{-\Lambda }\text{,}
\end{equation}%
which hold for an arbitrary $m>0$ ($C_{\infty }$ is a positive integration
constant). Here, we have defined
\begin{equation}
\Lambda =1+\sqrt{1+2h\left( gr_{0}\right) ^{2}}
\end{equation}%
as the parameter which controls the way the basic fields approach the
boundary $r\rightarrow \infty $. It is interesting to note that the source
field gives rise to the very same asymptotic behavior also found in the
previous MCSH case, including the dependence of the parameter $\Lambda $ on
the values of $r_{0}$, see the Eq. (\ref{ww1}).
\begin{figure}[t]
\includegraphics[width=8.4cm]{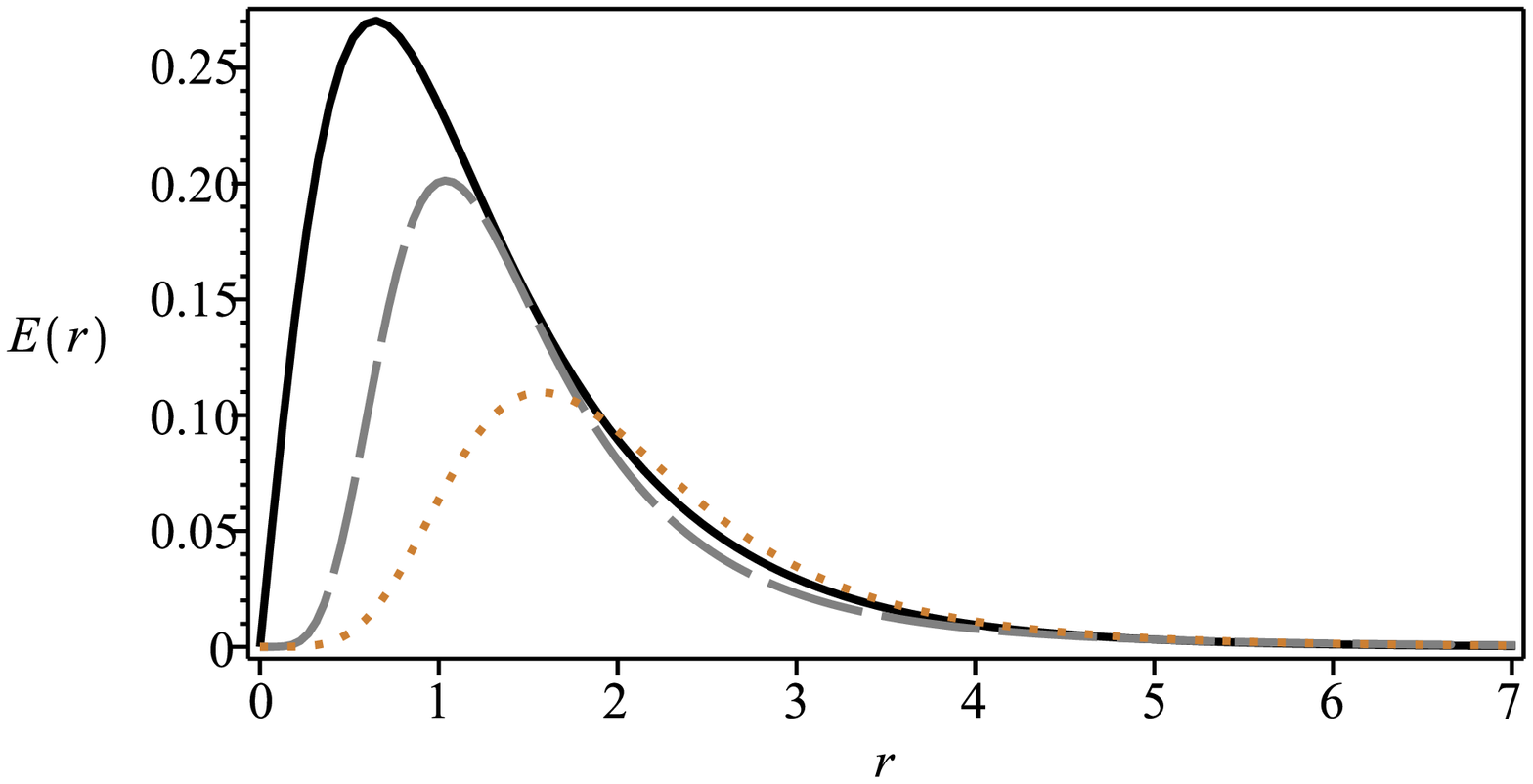} %
\includegraphics[width=8.4cm]{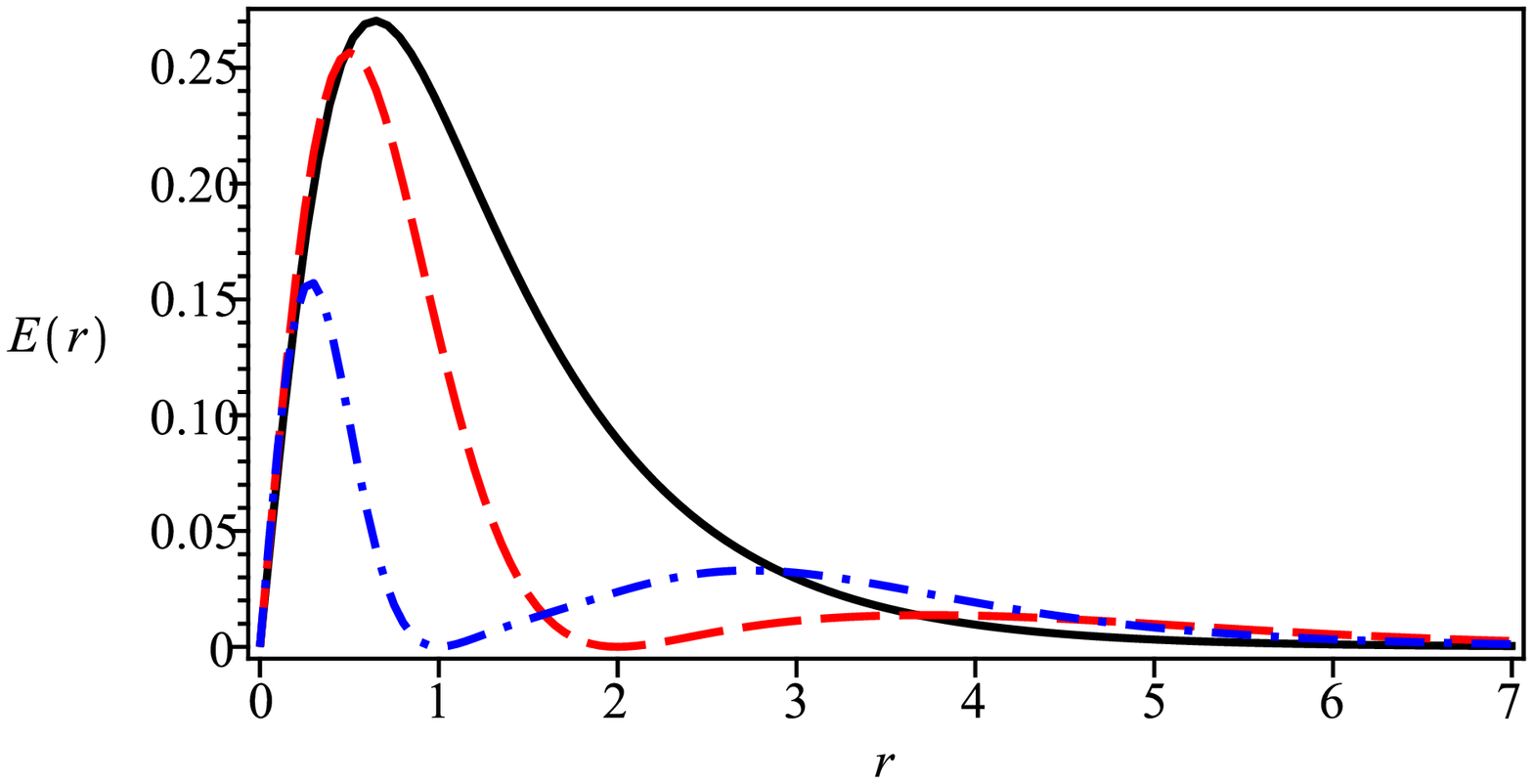}
\caption{Numerical solutions to the electric field $E(r)$. Conventions as in
the \ref{figg7}. The solutions for the 2nd. case vanish at $r=r_{0}$ in order to
avoid the second term in the right-hand side of the Eq. (\protect\ref{ed1})
to be singular.}
\end{figure}

\subsubsection{The second case}

We finally study the effects induced on the MCS-$CP(2)$ solitons by the
second choice for $G(\chi )$ introduced in the Eq. (\ref{tg2}) and
explicitly given in terms of $r$ by the Eq. (\ref{tg22}), i.e.%
\begin{equation}
G(\chi )=\frac{1}{\chi ^{2}}=\frac{\left( r^{2}+r_{0}^{2}\right) ^{2}}{%
\left( r^{2}-r_{0}^{2}\right) ^{2}}\text{.}
\end{equation}%
As already commented in the context of the MCSH case, its influence at the
boundaries disappears, from which the behavior of the corresponding solitons
is expected to mimic the one which emerges in the absence of the dielectric
medium. Furthermore, the electric and magnetic fields vanish at $r=r_{0}$,
then the first two terms in the right-hand side of the Eq. (\ref{ed1}) are
expected to be nonsingular.

In this case, the first-order equations (\ref{bpsx1}) and (\ref{bpsx2}) are
written, respectively, as%
\begin{equation}
\frac{1}{r}\frac{dA}{dr}=\frac{\left( r^{2}-r_{0}^{2}\right) ^{2}}{\left(
r^{2}+r_{0}^{2}\right) ^{2}}\left( \pm hg^{2}\cos \alpha -g\kappa
A_{0}\right) \text{,}  \label{y1}
\end{equation}%
\begin{equation}
\frac{d\alpha }{dr}=\pm \frac{\left( 2m-A\right) }{2r}\sin \alpha \text{,}
\label{y2}
\end{equation}%
while the Gauss law reads%
\begin{equation}
\frac{1}{r}\frac{d}{dr}\left[ \frac{\left( r^{2}+r_{0}^{2}\right) ^{2}r}{%
\left( r^{2}-r_{0}^{2}\right) ^{2}}\frac{dA_{0}}{dr}\right] +\kappa B=\frac{%
g^{2}h}{2}A_{0}\sin ^{2}\alpha \text{,}  \label{y3}
\end{equation}%
where the magnetic field $B$ is again given by the Eq. (\ref{Bcp2}).
\begin{figure}[t]
\includegraphics[width=8.4cm]{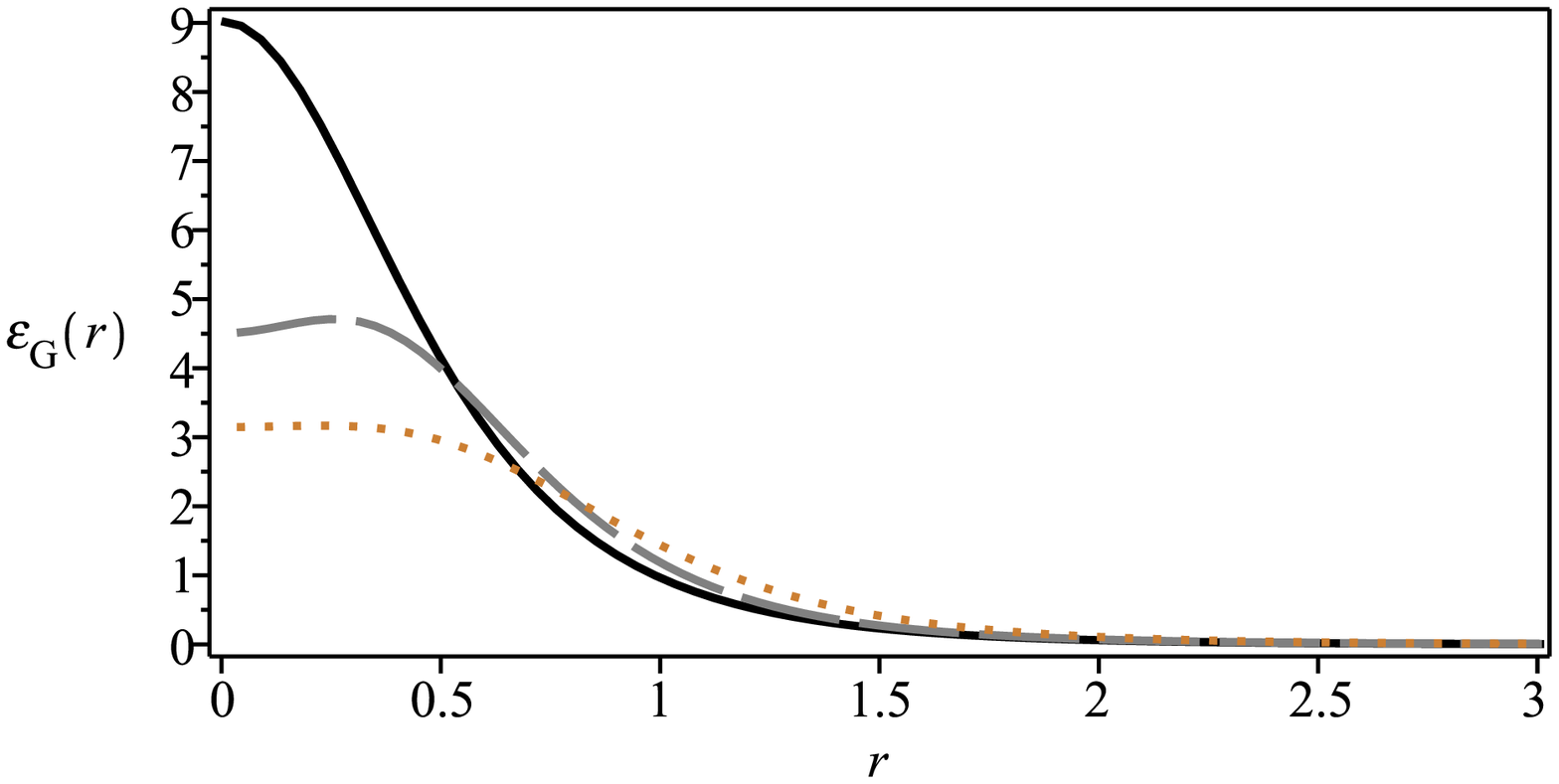} %
\includegraphics[width=8.4cm]{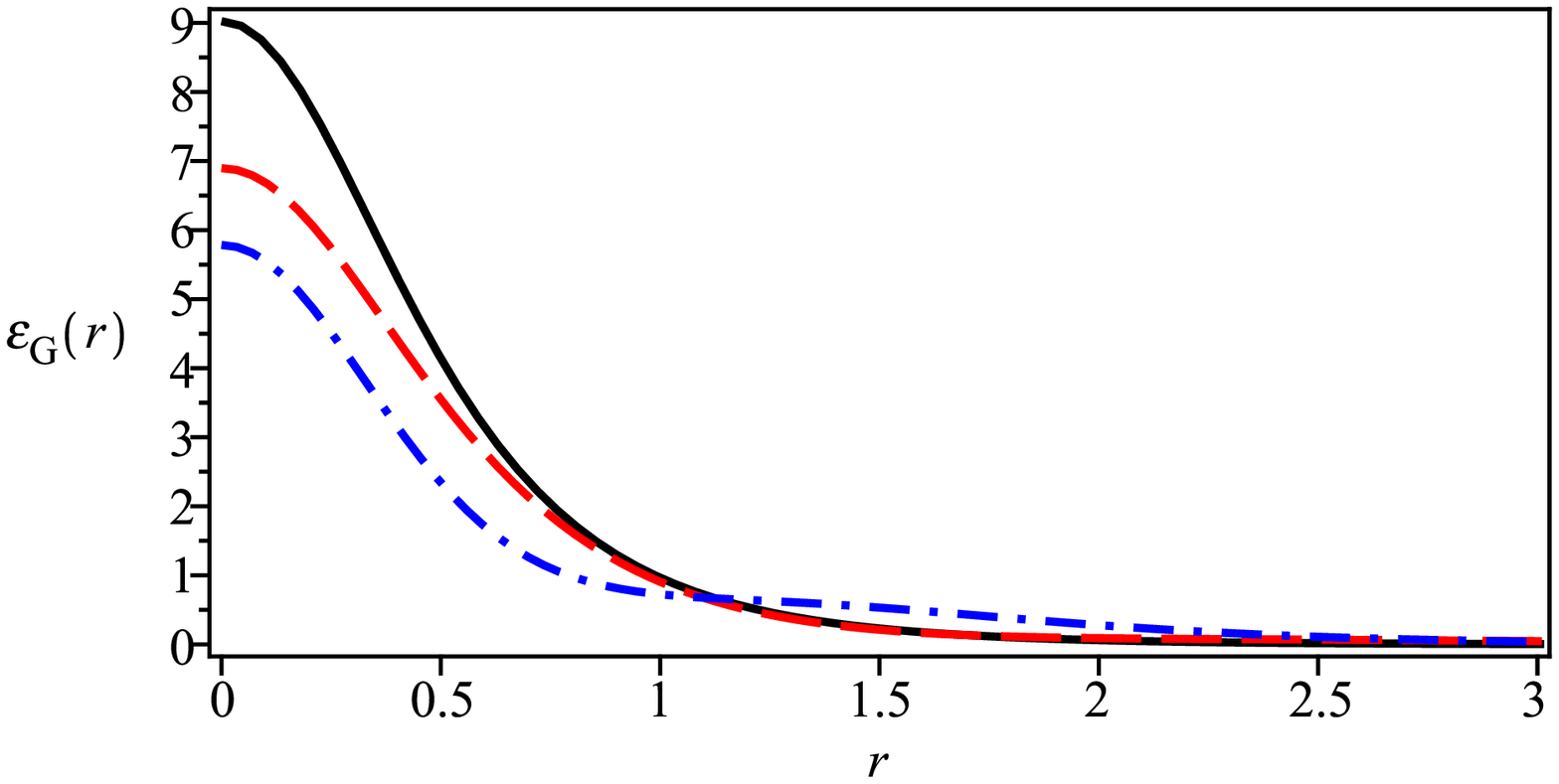}
\caption{Numerical solutions to the energy distribution $\protect\varepsilon %
_{G}$ related to the BPS configurations. Conventions as in the Fig. \ref{figg7}. The source field eventually changes the shape of the resulting profiles.}
\end{figure}

With the aim to compare the new results with the ones achieved previously,
we again set $h=\kappa =1$, $g=2$ and $m=1$, from which we solve the
differential equations (\ref{y1}), (\ref{y2}) and (\ref{y3}) for $r_{0}=1$
(dash-dotted blue line) and $r_{0}=2$ (dashed red line) according the
conditions (\ref{m4}), (\ref{m5}) and (\ref{b3}). We plot the resulting
profiles in the corresponding figures 7-12.

The resulting profiles for $\alpha (r)$, $A(r)$ and $A_{0}(r)$ appear in the
figures 7, 8, and 9, respectively, via which we see that again both the
gauge profile function and the electric potential present the formation of a
plateau in the neighborhood of $r=r_{0}$. As the reader can expect based on
our previous discussions, these structures force both the magnetic and
electric field to vanish at this particular point (see the figures 10 and
11, respectively), a similar effect to the one observed previously in the
MCSH case. Thus, they guarantee that the energy of the first-order
configurations converges to the value given by the Eq. (\ref{te2}).

The Figure 12 brings the profiles which we have found for the energy density
$\varepsilon_{G}$, from which we see that even in the presence of the source
field the new solutions mimic the same general shape presented by the
canonical configurations.

We end this study by investigating the behavior of the solutions near the
boundaries. In this sense, by solving the equations (\ref{y1}), (\ref{y2})
and (\ref{y3}) around to the origin, we obtain that the fields behave as%
\begin{equation}
\alpha (r)\approx \alpha _{1}r-\left[ \frac{\left( \alpha _{1}\right) ^{3}}{%
12}+\frac{g\tilde{B}_{0}}{8}\alpha _{1}\right] r^{3}\text{,}
\end{equation}%
\begin{equation}
A(r)\approx \frac{g\tilde{B}_{0}}{2}r^{2}-\tilde{A}_{1}r^{4}\text{,}
\end{equation}%
\begin{equation}
A_{0}(r)\approx w_{0}-\frac{\kappa \tilde{B}_{0}}{4}r^{2}\text{,}
\end{equation}%
with the expressions above holding for $m=1$, the winding number analyzed in
this work. The quantity $\tilde{B}_{0}$ was\ already\ defined in the Eq. (%
\ref{BBcp2}). Here, we have also introduced the parameter $\tilde{A}_{1}$
defined by%
\begin{equation}
\tilde{A}_{1}=\frac{g^{2}h}{8}(\alpha _{1})^{2}-\frac{g\tilde{B}_{0}}{16}%
\left( \kappa ^{2}-\frac{16}{r_{0}^{2}}\right) \text{.}
\end{equation}

The approximate solutions for $r\rightarrow \infty $ read as%
\begin{equation}
\alpha (r)\approx \frac{\pi }{2}-C_{\infty }\frac{e^{-Mr}}{\sqrt{r}}\text{,}
\end{equation}%
\begin{equation}
A(r)\approx 2m-2MC_{\infty }\sqrt{r}\,e^{-Mr}\text{,}
\end{equation}%
\begin{equation}
A_{0}(r)\approx \frac{2MC_{\infty }}{g}\frac{e^{-Mr}}{\sqrt{r}}\text{,}
\end{equation}%
which are valid for $m>0$. In this case, we have defined%
\begin{equation}
M=\frac{1}{2}\sqrt{2hg^{2}+\kappa ^{2}}-\frac{\left\vert \kappa \right\vert
}{2}\text{,}
\end{equation}%
which stands for the mass of the corresponding bosons. In view of these
asymptotic solutions, we conclude that, as in the previous MCSH scenario,
the presence of the dielectric medium does not change the way the basic
fields approach their boundary values, at least at relevant orders of $r$.

%%%%%%%%%%%%%%%%%%%%%%%%

\section{Summary and perspectives}

We have investigated the formation of BPS vortices with internal structures
in both the Maxwell-Chern-Simons-Higgs and the Maxwell-Chern-Simons-$CP(2)$
scenarios. The internal structures are generated through the introduction of
a dielectric medium which is driven by an extra real scalar field (named the
source field) which extends both the original models. We have then focused
our attention on time-independent radially symmetric configurations with a
self-dual structure. In order to attain such a goal, we have proceeded with
the implementation of the Bogomolnyi-Prasad-Sommerfield algorithm which has
allowed us to obtain the Bogomol'nyi bound and the self-dual BPS equations
for both the models analyzed in this manuscript.

In the sequence, we have observed that the self-dual equation for the source
field depends only on the particular choice for the superpotential $W(\chi )$
and that therefore this equation decouples from the others BPS equations of
the model. For a specific superpotential, the self-dual solution for $\chi
(r)$ allows us to define a dielectric function $G(\chi )$ which becomes
responsible for the generation of the internal structures inherent to the
new first-order vortices. We have splitted our investigation into two
different branches based on the functional forms which we have chosen for
the dielectric function $G(\chi )$ which appears in the Lagrange densities
of the models.

We have finally solved the remaining BPS equations together with the Gauss
law by means of a finite-difference scheme according to the canonical
boundary conditions. The numerical results have verified that the presence
of the dielectric medium induces similar effects in both scenarios.

In view of the numerical analysis, we have observed that the dielectric
causes different effects on the corresponding vortex profiles in comparison
to the ones obtained in the absence of if. This way, the first choice $%
G(\chi )=(1-\chi ^{2})^{-1}$ changes the way the magnetic and electric
fields behave near the boundaries. On the other hand, the second choice $%
G(\chi )=\chi ^{-2}$ affects the manner the fields behave along the radial
coordinate, except near the boundary values. In particular, the second
choice for the dielectric function leads to magnetic and electric fields
which vanish at $r=r_{0}$. This particular behavior ensures that the energy
of the self-dual BPS vortices converge to the finite Bogomol'nyi bound.

An interesting perspective is the application of the present idea to the
Maxwell-Skyrme system, or the study of the MCS-$CP(2)$ vortices with an
unusual shape caused by the inclusion of a magnetic impurity, see the
references \cite{15} and \cite{19}, for instance. These issues are currently
under investigation and the results we will be reported in a future
contribution.

\bigskip \medskip

\begin{acknowledgments}
This study was financed in part by the Coordena\c{c}\~{a}o de Aperfei\c{c}%
oamento de Pessoal de N\'{\i}vel Superior - Brasil (CAPES) - Finance Code
001, the Conselho Nacional de Desenvolvimento Cient\'{\i}fico e Tecnol\'{o}%
gico - Brasil (CNPq) and the Funda\c{c}\~{a}o de Amparo \`{a} Pesquisa e ao
Desenvolvimento Cient\'{\i}fico e Tecnol\'{o}gico do Maranh\~{a}o - Brasil
(FAPEMA). J. A. thanks the full support from CAPES. R. C.  acknowledges the support from the grants CNPq/306724/2019-7, CNPq/423862/2018-9 and FAPEMA/Universal-01131/17. E. H. thanks the support from the grants
CNPq/307545/2016-4 and FAPEMA/COOPI/07838/17. EH also acknowledges the
School of Mathematics, Statistics and Actuarial Science of the University of
Kent (Canterbury, United Kingdom) for the kind hospitality during the
realization of part of this work.
\end{acknowledgments}

\end{document}